\pdfoutput=1
\documentclass[11pt]{article}
\usepackage{amssymb}
\usepackage{graphicx}

\parskip \baselineskip

\newcommand{\ds}{\displaystyle}

\newcommand{\ol}{\overline}

\newcommand{\ra}{\rightarrow}
\newcommand{\Ra}{\Rightarrow}

\begin{document}
\title{Compressed representation of Learning Spaces}

\author{Marcel Wild\\ Department of Mathematical Sciences,\\ University of Stellenbosch,\\
Matieland, 7600\\ South Africa\\
\texttt{mwild@sun.ac.za}}

\date{}
\maketitle

\begin{abstract}
Learning Spaces are certain set systems that are applied in the mathematical modeling of education. We propose a wildcard-based compression (without loss of information) of such set systems to facilitate their logical and statistical analysis. Under certain circumstances compression is the prerequisite to calculate the Learning Space in the first place.  There are connections to the dual framework of Formal Concept Analysis and in particular to so called attribute exploration.
\end{abstract}


\section{Introduction}

In order to grasp the structure of this article one needs a basic understanding of what Knowledge Spaces and the more specific Learning Spaces are all about. Nothwithstanding initial concerns of the Referees we begin with a long verbatim quotation of [D]. It makes up the whole of Subsection 1.1 and, in the author's opinion, is the perfect way to introduce Learning Spaces to the novice. Only afterwards we will be in a position to state our main contribution (in 1.2), and to proceed with the Section break up (in 1.3).

{\bf 1.1} In Knowledge Space Theory (KST) a `knowledge structure' encodes a body of information as a `domain' together with `states of knowledge'. The domain is the set of all the relevant, elementary pieces of information. Each knowledge state is a subset of the domain, which contains all the items mastered at some time by some (hypothetical) individual. For example, the empty set and the domain itself represent respectively a completely ignorant and an omniscient student. In general, there will be many more knowledge states; their collection captures the overall structure of the body of information. If $Q$ is the domain and ${\cal K}$ the collection of states, the knowledge structure is the pair $(Q,{\cal K})$. An example with domain $Q_1 = \{a, b, c, d\}$ is displayed in Figure 1: the boxes show the nine states forming ${\cal K}_1$, while the ascending lines indicate the covering relation among states.

\begin{center}
\includegraphics{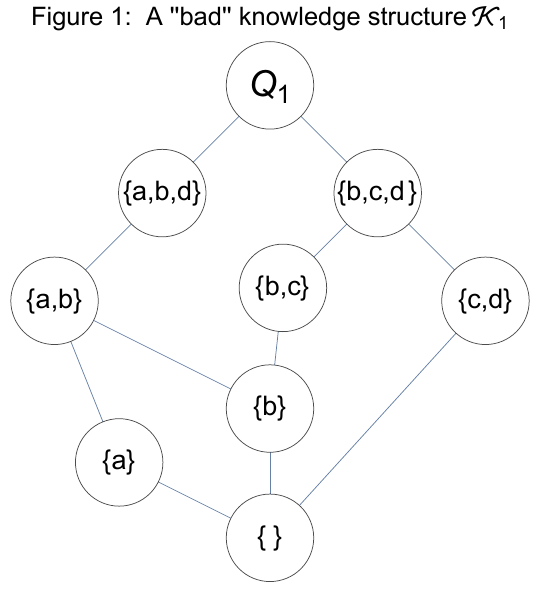}
\end{center}

Without further restrictions on the collection of states, knowledge structures are too poorly organized for the development of a useful theory. Fortunately, pedagogical considerations lead in a natural way to impose restrictions on the state collection. We now explain two natural requirements by looking at the knowledge structure $(Q_1, {\cal K}_1)$ from Figure 1. The subset $\{c, d\}$ is a knowledge state in ${\cal K}_1$, but there is no way for a student to acquire mastery of items $c$ and $d$ one after the other in any order (neither subset $\{c\}$ nor $\{d\}$ is a state in ${\cal K}_1$). This contradicts the (common) view that learning occurs progressively, that is one item at a time. For another singularity in the same knowledge structure $(Q_1, {\cal K}_1)$, consider a student in state $\{b\}$. She may learn item $a$ to reach state $\{a, b\}$. On the other hand, while in state $\{b\}$ she may rather learn item $c$ first and reach state $\{b, c\}$; then, strangely enough, item $a$ is not learnable anymore to her (because the subset $\{a, b, c\}$ is not a state in ${\cal K}_1$). The definiton of a `Learning Space' as a  particular type of knowledge structure rules out the two strange situations that we just illustrated on Figure 1. It imposes the following two conditions on the  states of a knowledge structure $(Q, {\cal K})$.
\begin{enumerate}
	\item[{(A)}] ACCESSIBILITY. Any state $K$ contains an item $q$ such that $K \setminus \{q\}$ is again a state.
	\item[{(LC)}] LEARNING CONSISTENCY. For a state $K$ and items $q, r$ if $K \cup \{q\}$ and $K \cup \{r\}$ are states, then $K \cup \{q, r\}$ is also a state. 
\end{enumerate}
{\bf 1.2} The primary purpose of this article is the application of compression techniques (previously explored by the author in other contexts) to accomodate knowledge structures with millions of states. That not only reduces storage space but facilitates statistical analysis.  In the framework of the more specific Learning Spaces these compression techniques naturally lead to ``query learning'' which constitutes the second theme of our article. Although the author's expertise is skewed towards the first theme, the research directions proposed for the second are deemed to be fruitful. All in all, the present article is heavier on mathematics and algorithms than the average article in this journal.

{\bf 1.3} Here comes the section break up. Section 2 introduces, by way of a toy example, the basic idea  of how large chunks of the powerset ${\cal P}(Q)$ can be chopped away in such a way that the desired knowledge structure ${\cal K}$ results in a compressed fashion. 

Section 3 presents both well and lesser known facts about specific knowledge structures, i.e. so called Knowledge Spaces $(Q,{\cal K})$. In 3.1 we introduce the base ${\cal B}({\cal K}) \subseteq {\cal K}$. This leads (3.2) to Dowling's algorithm that generates ${\cal K}$ from ${\cal B}({\cal K})$. In 3.3 we make precise the informal ``dual implications'' ($=$ dimplications) occuring in Section 2. (The matching term in [FD] is ``entailment''.) Of particular importance are {\it prime} dimplications. In Theorem 1 we show how the set PrimeDimp$({\cal K})$ of all prime dimplications of a Knowledge Space ${\cal K}$ can be calculated from ${\cal B}({\cal K})$. In 3.4 we see that PrimeDimp$({\cal K})$ is just one example (though an important one) of a ``dimplication base'' of ${\cal K}$. Subsection 3.5 is about Learning Spaces ${\cal K}$, as defined by (A) and (LC) above. Learning Spaces are Knowledge Spaces ${\cal K}$ for which both ${\cal B}({\cal K})$ and PrimeDimp$({\cal K})$ are particularly well behaved. Subsection 3.6 points out that Learning Spaces are known as antimatroids in the Combinatorics and Operations Research communities.

Section 4 is in the spirit of Section 2 but with more sophisticated don't-care symbols (aka wildcards). The underlying $e${\it  -algorithm} was previously applied by the author in other circumstances. Here we show that, given any base $\Theta$ of dimplications of an (unknown) Knowledge Space ${\cal K}$, the $e$-algorithm can calculate a compact representation of ${\cal K}$. In 4.2 the latter is used for statistical analysis (as alluded to in 1.2), and in 4.3 we show how the base ${\cal B}({\cal K})$ can be sieved from it.

Section 5 recalls the duality between Knowledge Spaces ${\cal K}$ and Closure Spaces ${\cal C}$. In particular, dimplications correspond to the better known implications. In 5.1 we introduce lattices and show how each lattice ${\cal L}$ can be modelled  naturally by a closure system ${\cal C} ({\cal L})$. This allows to apply the theory of implications to lattices. Many specific lattices have been investigated in this regard, see [W2] for a survey. For our purpose so called meet-semidistributive lattices come into focus; the relevant facts are readied in 5.2.

This is exploited in Section 6 where it leads to a second method to compress a Learning Space, apart from the way in Section 4 which works for any Knowledge Space. In brief, whereas the $e$-algorithm from Section 4 operates on the universe $Q$, the $n$-algorithm from Section 6 has ${\cal B}$ as its universe, and always $|{\cal B}| \geq |Q|$. As opposed to $\Theta$ in Section 4, the size of the base $\Sigma$ of implications derived from ${\cal B}$ is $|\Sigma| \leq |{\cal B}|^2$ by Theorem 2.
Section 7 evaluates the discussed algorithms on computer-generated random examples. Section 8 dwells on the ``query learning'' aspect of it all. The framework of Formal Concept Analysis, will be compared to Knowledge Space Theory, and we glimpse at the general theory of learning Boolean functions.

\section{Compression of knowledge structures using don't-care symbols}

A dual implication or briefly ``dimplication'' is a certain statement about a knowledge structure which is either true or false. To fix ideas, consider the knowledge structure ${\cal K}_2$ in Figure 2 (which is based on Fig.15.1 in [FD]) with domain $Q_2 = \{a, b, c, d,e\}$. By definition the dimplication $\{b, d\} \rightsquigarrow c$ ``holds'' in ${\cal K}_2$ when every student who fails both $b$ and $d$ also fails $c$. Put another way: The mastering of $c$ implies the mastering of $b$ or $d$. If ${\cal K}_2$ is {\it known} in one way or another, e.g. in diagram form as in Figure 2, then it is easy in principle (but possibly tiresome in practise) to decide whether some dimplication holds. In our case $\{b, d\} \rightsquigarrow c$ holds in ${\cal K}_2$ because (check) every knowledge state $K \in {\cal K}_2$ that contains $c$ also contains $b$ or $d$. Likewise $\{b,c\} \rightsquigarrow e$ does {\it not} hold in ${\cal K}_2$ because (say) $K = \{a, e\}$ contains $e$ but neither $b$ nor $c$.

{\bf 2.1} Before continuing with our toy example ${\cal K}_2$ it pays to properly formalize dimplications. This concept was introduced in [Ko] as ``entailment'' (see also [FD, p.44]) but we give it another name in order to better match the established terminology of Section 5. Thus if $A, B$ are {\it nonvoid} and {\it disjoint} subsets of some fixed set $Q$ (the ``domain'')  then the expression $A \rightsquigarrow B$ is called a {\it dimplication}. It {\it holds} for a {\it subset} $S \subseteq Q$ (or : $S$ {\it satisfies} $A \rightsquigarrow B$) if

(1) \quad $A \cap S = \emptyset \ \Ra \ B\cap S = \emptyset$.

There are two ways for $A \rightsquigarrow B$ to hold in $S$: Either $A \cap S = \emptyset$ and thus $B \cap S = \emptyset$. Or $A \cap S \neq \emptyset$, in which case there is no further requirement. (One could cut the cake another way, but we only do that this one time: Either $B \cap S \neq \emptyset$ and thus $A \cap S \neq \emptyset$. Or $B \cap S = \emptyset$, in which case there is no further requirement.) We say that $A \rightsquigarrow B$ {\it holds for the knowledge structure} ${\cal K}$ if $A \rightsquigarrow B$ holds for each $S \in {\cal K}$. Say $B = \{b_1, \cdots, b_s\}$. Then one checks that $A \rightsquigarrow B$ holds for ${\cal K}$ if and only if each $A \rightsquigarrow \{b_i\}$ holds for ${\cal K} \ (1 \leq i \leq s)$. We shall often deal with dimplications $A \rightsquigarrow \{b\}$ in the first place and sometimes write $A \rightsquigarrow b$ instead of $A \rightsquigarrow \{b\}$.

{\bf 2.2} Suppose the knowledge structure ${\cal K}$ is {\it not known} to us, but there is an {\it expert} available insofar that she can answer correctly\footnote{For this the expert dosen't need to have a diagram of ${\cal K}$ in front of her. It could be that the dimplication is provable in a strictly logical sense, akin to the affirmative answer of ``When 17 divides $ab$, does $17$ divide $a$ or $b$?'' For other types of dimplications we trust her ``gut feeling''. In such a situation 100\% accuracy of the expert is  unlikely, but we imagine it for the sake of argument. (Even 90\% accuracy would still yield good results.) More on that in Section 8.} whether or not any proposed dimplication holds. Before we ask the first question the sought knowledge structure ${\cal K}$ potentially  equals ${\cal F} = {\cal P}(Q)$, i.e. the powerset of the domain $Q$ of ${\cal K}$. 

More generally, suppose that we have advanced to a collection ${\cal F} \subseteq {\cal P}(Q)$ of remaining potential states and that we then get a positive answer to the dimplication $A \rightsquigarrow b$. The latter rules out the subsets in ${\cal F}$ which are disjoint from $A$ yet contain $b$. The remaining subsets of ${\cal F}$, i.e. the ones satisfying $A \rightsquigarrow b$, fall in two disjoint subcollections of ${\cal F}$, say ${\cal F}_1$ and ${\cal F}_2$. Namely, ${\cal F}_1$ contains all sets of ${\cal F}$ disjoint from $A$ and (whence) not containing $b$ either. And ${\cal F}_2$ contains all sets of ${\cal F}$ that intersect $A$. We thus need to shrink the old collection ${\cal F}$ to ${\cal F}_1 \uplus {\cal F}_2$. Here $\uplus$ denotes disjoint union. 

When we ask about the validity of $A \rightsquigarrow b$ and get a negative answer from the expert then no\footnote{Although negative answers provide information as well, our kind of algorithm can only react upon positive answers.} action needs to be taken. 

{\bf 2.3} To fix ideas, let ${\cal K} = {\cal K}_2$ from before and suppose that among the list of dimplications that we queried, the ones that received a positive answer were:

(2) \quad $\{e\} \rightsquigarrow a, \ \ \{a\} \rightsquigarrow b, \ \ \{b, d\} \rightsquigarrow c$.

We start by writing the powerset of $Q_2 = \{a, b, c, d, e\}$ as $(2,2,2,2,2)$ with the understanding that each don't-care symbol ``2'' is free to be $0$ or $1$. For instance, the seemingly strange expression $(1,0,1,0,1) \in (2,2,2,2,2)$ makes perfect sense. It just corresponds (while being more handy) to the formula $\{a, c, e\} \in {\cal P} (Q_2)$. Starting with the first member $\{e\} \rightsquigarrow a$ in (2) we need to shrink ${\cal F} : = (2,2,2,2,2)$ to ${\cal F}_1 \uplus {\cal F}_2$ as explained above. Hence ${\cal F}_1$ consists of all $S \in {\cal F}$ with $S \cap \{e\} = \emptyset$ and $a \not\in S$, while ${\cal F}_2$ consists of all $S \in {\cal F}$ with $S \cap \{e\} \neq \emptyset$. Evidently ${\cal F}_1$ and ${\cal F}_2$ can be encoded, respectively,  by the $012${\it -rows} $r_1$ and $r_2$ in Table 1. Our new ${\cal F}$ is now ${\cal F}: = r_1 \uplus r_2$.

According to (2) the next dimplication to be {\it imposed} in this  way is $\{a\} \rightsquigarrow b$. The family of sets $S \in r_1$ satisfying $\{a\} \rightsquigarrow b$ obviously is $r_3$. One could be tempted to similarly impose $\{a\} \rightsquigarrow b$ upon $r_2$ right away. However, it is a better and well established strategy (more on that in 2.4) to always only process the {\it top} row of the {\it working stack}. Currently our working stack has two members, i.e. $r_3$ on top of $r_2$. In order to enable the ``top-row-strategy'' we need to keep track, for  each row of the working stack, which dimplication is {\it pending}. Currently $r_3$ has the third dimplication in (2) pending and $r_2$ the second. 

A moment's thought shows that imposing $\{b, d\} \rightsquigarrow c$ upon top row $r_3$ results in $r_3$ being replaced by $r_4 \uplus r_5$.  Rows $r_4$ and $r_5$ are {\it final} in the sense that all dimplications in list (2) have been imposed on them. Hence $r_4$ and $r_5$ are removed from the working stack and stored somewhere else. 

Thus the new top row (incidently the working stack's only row) is $r_2$. It is clear that imposing $r_2$'s pending dimplication $\{a \} \rightsquigarrow b$ on it replaces $r_2$ by $r_6 \uplus r_7$. Imposing $\{b, d\} \rightsquigarrow c$ upon the top row $r_6$ replaces $r_6$ by $r_8 \uplus r_9$. Again $r_8$ and $r_9$ are final and get removed. The new top row is $r_7$. In order to impose $\{b, d\} \rightsquigarrow c$ upon $r_7$ let us reactivate previous notation and put ${\cal F}_1: = \{S \in r_7: S \cap \{b , d\} = \emptyset \}$ and ${\cal F}_2 : = \{S \in r_7: S \cap \{b, d\} \neq \emptyset \}$. Obviously the sets in ${\cal F}_1$ that satisfy $\{b, d\}\rightsquigarrow c$ are the ones in $r_{10}$. Ditto the sets in ${\cal F}_2$ that satisfy $\{b, d\} \rightsquigarrow \{c\}$ are\footnote{Notice that $r_{11} \uplus r_{12} = r \cup r' : = (1, 1, 2, 2, 1) \cup (1,2,2,1,1)$ but other than $r_{11}, r_{12}$ the rows $r, r'$ are not disjoint; e.g. $(1,1,0,1,1) \in r_{11} \cap r_{12}$. In Section 4 we give more systematic ways to achieving disjointness.} the ones in $r_{11} \uplus r_{12}$.


\begin{center}
\includegraphics{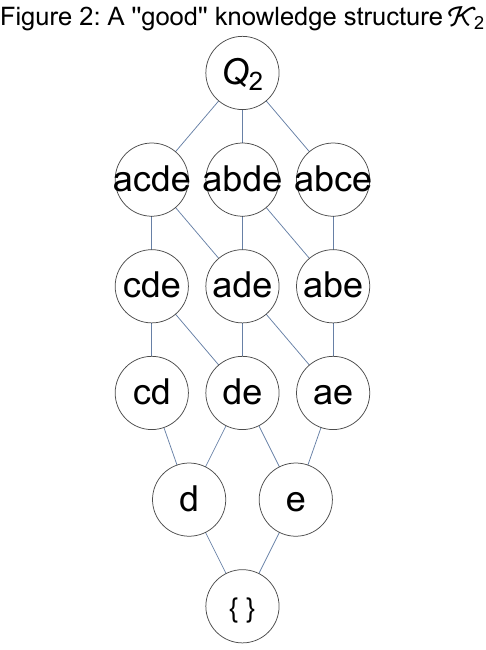}
\end{center}

\begin{tabular}{l|c|c|c|c|c|l} 
& $a$ & $b$ & $c$ & $d$ & $e$ & \\ \hline
&  &   &   &  &  &\\ \hline
$r_1=$ & 0 & 2 & 2 & 2 & 0 &\\
$r_2=$ & 2 & 2 & 2 & 2 & 1 &\\ \hline
 &  &  &   &  & & \\ \hline 
 $r_3=$ & 0 & 0 & 2 & 2 & 0 & pending dimp. 3\\
 $r_2 =$ & 2 & 2 & 2 & 2 & 1 & pending dimp. 2\\ \hline
 &   &    &  &  & & \\ \hline
 $r_4=$ & 0 &0 & 0 & ${\bf 0}$ & 0 & final \\
 $r_5 =$ & 0 & 0 & 2 & ${\bf 1}$ & 0 & final \\
 $r_2 =$ & 2 & 2 & 2 & 2 & 1 & pending dimp. 2\\ \hline 
 & & & & & & \\ \hline 
 $r_6 =$ & ${\bf 0}$ & 0 & 2 & 2 & 1 & pending dimp. 3 \\
 $r_7 =$ & ${\bf 1}$ & 2 & 2 & 2 & 1 & pending dimp. 3 \\ \hline
 & & & & & & \\ \hline
 $r_8=$ & 0 & 0 & 0 & ${\bf 0}$ & 1 & final \\ \hline
 $r_9=$ &0 & 0 & 2 & ${\bf 1}$ & 1 & final\\ \hline
 $r_7=$ & 1 & 2 & 2 & 2 & 1 & pending dimp. 3\\ \hline
 & & & & & & \\ \hline
 $r_{10} =$ & 1 & 0 &0 & 0 & 1 & final \\ \hline 
 $r_{11}=$ & 1 & ${\bf 1}$ & 2 & ${\bf 2}$ & 1 & final\\ \hline
 $r_{12}=$ & 1 & ${\bf 0}$ & 2 & ${\bf 1}$ & 1 & final \\ \hline
 \end{tabular}
 
 Table 1: Compressing a knowledge structure with $012$-rows
 
 One verifies that the union of final rows $r_4 \cup r_5 \cup r_8 \cup r_9 \cup r_{10} \cup r_{11} \cup r_{12}$ coincides with the knowledge structure $(Q_2, {\cal K}_2)$ in Figure 2. For instance $\{a, d, e\}$ in Figure 2 is a member of $r_{12}$. Suffice it to check that the cardinalities match:
 $$|r_4| + |r_5| + \cdots + |r_{12}| = 1+ 2 + 1 + 2 + 1 + 4 + 2 = 13.$$

{\bf 2.4} The strategy to always process the top row of a (working) stack is well known among computer scientists and goes under the name ``Last In, First Out'' (LIFO).

\begin{center}
\includegraphics{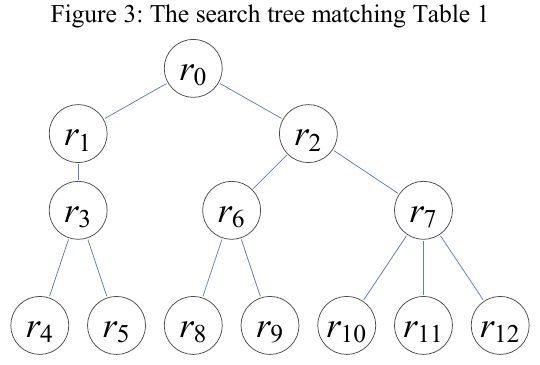}
\end{center}

As is well known, each LIFO strategy amounts to a so-called depth-first search of a tree. In our case the tree is depicted in Figure 3 (where $r_0 : = (2, 2, 2, 2, 2)$). Other than usual for depth-first searches, here the {\it leaves} $r_4, r_5, r_8, r_9, r_{10}, r_{11}, r_{12}$ do not correspond to individual models, but to {\it sets of models}. For instance
$$r_{11} = \{\{a, b, e\}, \{a, b, c, e\}, \{a, b, d, e\}, \{a, b, c, d, e\}\}$$

\section{On the mathematics of Knowledge Spaces}

Coming back to the knowledge structure $(Q_1, {\cal K}_1)$ from Figure 1, note that the knowledge states $\{a, b\}$ and $\{b, c\}$ belong to ${\cal K}_1$, yet their union $\{a, b, c\}$ doesn't. A knowledge structure $(Q, {\cal K})$ which avoids this type of anomality is called a {\it Knowledge Space}\footnote{We mention in passing that evidently each Knowledge Space satisfies (LC), but the converse fails.}. Thus it holds that

(3) \quad $K \cup L \in {\cal K} \ \mbox{for all} \ K, L \in {\cal K}$.

To avoid trivial cases we further postulate that $\emptyset, Q \in {\cal K}$. As a consequence of (3), each subset $S \subseteq Q$ contains a largest knowledge state, the so-called {\it interior} of $S$, which is defined as

(4) \quad $S^0 : = \bigcup \{K \in {\cal K} | \ K \subseteq S \}.$

We now discuss two ways to grasp Knowledge Spaces; the first (3.1 and 3.2) is the base  ${\cal B} \subseteq {\cal K}$, the second (3.3 and 3.4) the use of dimplications. In 3.5 and 3.6 we look at the peculiarities occuring for special types of Knowledge Spaces, i.e. Learning Spaces.

{\bf 3.1} For all details left out in this subsection see [FD, ch.3.4]. Let $(Q, {\cal K})$ be a Knowledge Space. For any element $m \in Q$ an {\it atom at $m$} is a minimal set $P \in {\cal K}$ containing $m$. The minimality condition readily implies that $P$ is {\it join-irreducible} in the sense that $P$ cannot be the union of knowledge states strictly contained in $P$. We define:
$$\begin{array}{lll}
\mbox{Atoms}(m) & : = & \{P \in {\cal K} : P \ \mbox{is atom at} \ m \} \ \ \ (m \in Q) \\
\\
{\cal B} = {\cal B}({\cal K})& := & \ds\bigcup_{m \in Q} \ \mbox{Atoms}(m) \end{array}$$
It turns out that ${\cal B}$ is the family of {\it all} join-irreducible sets of ${\cal K}$; we refer to ${\cal B}$ as the {\it base} of ${\cal K}$ and to its members as {\it base sets}. Each $K \in{\cal K}$ is a union of base sets, usually in many ways. For instance, writing e.g. $bc$ for $\{b, c\}$ the base of ${\cal K}_1$ in Figure 1 is ${\cal B} = \{a, b, bc, cd, abd\}$ and say Atoms$(c) = \{bc, cd\}$, Atoms$(d) = \{abd, cd\}$.

{\bf 3.2} Suppose the base\footnote{Actually here ${\cal B}$ could be {\it any} family of subsets such that each $K \in {\cal K}$ is a union of members of ${\cal B}$.} ${\cal B}$ of some (unknown) Knowledge Space ${\cal K} \subseteq {\cal P}(Q)$ is given. Then a natural way to generate ${\cal K}$ from ${\cal B} = \{B_1, B_2, \cdots, B_n\}$ is as follows. By induction let $K_1$ to $K_m$ be all (different) unions of sets from $\{B_1, \cdots,  B_{t-1}\}$. To handle $B_t$ add all sets $K_1 \cup B_t, K_2 \cup B_t, \cdots, K_m \cup B_t$ to the list. Trouble is, many sets $K_i \cup B_t$ may not be new, i.e. belong to $\{K_1, \cdots, K_m\}$. What is more, whether new or not,  $K_i \cup B_t = K_j \cup B_t$ for $i \neq j$ is possible. However, due to Dowling [Dow] (or see [FD, 3.53]) there is an efficient way to identify K$_{i_1}, K_{i_2}, \cdots, K_{i_\alpha}$ within $\{K_1, \cdots, K_m\}$ which avoid such duplications and satisfy

(5) \quad $\{K_{i_1} \cup B_t, \cdots, K_{i_\alpha} \cup B_t\} = \{K_1 \cup B_t, \cdots, K_m \cup B_t\}$.




{\bf 3.3} A dimplication $A \rightsquigarrow b$ holding in a Knowledge Space $(Q, {\cal K})$ is called a {\it prime} dimplication if $A_0 \rightsquigarrow b$ no longer holds when $A_0 \varsubsetneqq A$. For instance all of $e \rightsquigarrow a, \ a \rightsquigarrow b$, and $bd \rightsquigarrow c$ in Section 2 are prime, but e.g. $de \rightsquigarrow b$ is not since $e \rightsquigarrow b$ also holds. Let PrimeDimp$({\cal K})$ be the set of all prime dimplications of ${\cal K}$. 
In order to see how PrimeDimp$({\cal K})$ can be calculated from ${\cal B}$ we need some preliminaries. A {\it transversal} of a nonempty set system ${\cal S}$ is any set $T$ such that $T \cap X \neq \emptyset$ for all $X \in {\cal S}$. The transversal $T$ is {\it minimal} if no proper subset of $T$ is a transversal of ${\cal S}$. We denote by mintr$({\cal S})$ the family of all minimal transversals of ${\cal S}$. Note that $\emptyset \in {\cal S} \ \Leftrightarrow mintr ({\cal S}) = \emptyset$. Calculating $mintr({\cal S})$ has many applications, and a great variety of algorithms have been proposed; see [GV] for a nice survey. As to Theorem 1, it has been discovered independently in [GKL, Theorem 1.9], albeit in the special case of ``antimatroids'' (to be discussed in 3.6).

{\bf Theorem 1:} With notation as above and putting $Q': = \{q\in Q: \{q\} \not\in {\cal B}\}$ it holds that
PrimeDimp$({\cal K}) = \ds\bigcup_{b \in Q'} \{A \rightsquigarrow b: A \in MT(b)\},$
where $MT(b) : = \mbox{mintr}(\{P \setminus \{b\}: \ P \in Atoms(b) \})$.

Let us first investigate the extreme case $Q' = \emptyset$ (which nonetheless is covered by the proof below). So if $\{q\} \in B$ for all $q \in Q$, then clearly ${\cal B} = \{ \{q\}: q \in Q\}$. Thus ${\cal K}= {\cal P}(Q)$, and so {\it no} dimplication $A \rightsquigarrow b$ holds in ${\cal K}$. Indeed, $A \rightsquigarrow b$ e.g. fails in $(Q \setminus A) \in {\cal K}$. Consequently, if (and only if) ${\cal K} = {\cal P}(Q)$ then PrimeDimp$({\cal K}) = \emptyset$.

{\it Proof:} Notice first that $b \in Q'$ implies $\emptyset \not\in \{P \setminus \{b\}:  P \in Atoms(b) \}$ which implies $MT(b) \neq \emptyset$. Consider now any $A \rightsquigarrow b$ for fixed $b \in Q'$ and $A \in MT(b)$. To show that $A \rightsquigarrow b$ is a dimplication of $(Q, {\cal K})$ we take any $K \in {\cal K}$ and verify (1) for $B: = \{b\}$ and $S: = K$. Since there is nothing to show for $b \not\in K$, let $b \in K$. Then $b \in P \subseteq K$ for some $P \in \mbox{Atoms}(b)$. Since $A \cap (P \setminus \{b\}) \neq \emptyset$ in view of $A \in MT(b)$, one has $A \cap K \neq \emptyset$, and so (1) holds. To show that $A \rightsquigarrow b$ is a {\it prime} dimplication consider any $A _0 \rightsquigarrow b$ with $A_0 \varsubsetneqq A$. Since $A$ is a minimial transversal of $\{P \setminus \{b\}:  P \in Atoms(b)\}$ there is some $P_0 \in  \mbox{Atoms}(b)$ such that $b \in P_0$ and $A_0 \cap P_0 = \emptyset$. Since $A_0 \rightsquigarrow b$ does not hold in $P_0$, it follows that $A_0 \rightsquigarrow b$ is no dimplication of $(Q, {\cal K})$.

Conversely, let $A \rightsquigarrow b$ be any member of PrimeDimp$({\cal K})$. (Recall that $b \not\in A$ by convention in 2.1.) We need to show that $A \in MT(b)$. Since $A \rightsquigarrow b$ is a dimplication of $(Q, {\cal K})$, it holds for all $P \in {\cal K}$ with $b \in P$ that $A \cap P \neq \emptyset$. Because of $b \not\in A$ also $A \cap (P \setminus \{b\}) \neq \emptyset$. Thus $A$ is a transversal of ${\cal F}: = \{P \setminus \{b\}: P \in Atoms(b)\}$. 
Suppose by way of contradiction that $A$ was {\it no} minimal transversal of ${\cal F}$. Then there is $A_0 \varsubsetneqq A$ such that $A_0 \cap (P \setminus \{b\}) \neq \emptyset$ for all $P \setminus \{b\}$ in ${\cal F}$. In order to show that $A_0 \rightsquigarrow b$ holds in ${\cal K}$ (which is the desired contradiction to the primeness of $A \rightsquigarrow b$) consider any $K \in {\cal K}$ with $b \in K$. Then $b \in P_0 \subseteq K$ for some $P_0\in Atoms(b)$. From $A_0 \cap K \supseteq A_0 \cap (P_0 \setminus \{b\}) \neq \emptyset$ follows that $A_0 \rightsquigarrow b$ holds in $K$. Since $K$ was arbitrary, $A_0 \rightsquigarrow b$ holds in ${\cal K}$. \hfill $\square$

{\bf 3.4} Consider an arbitrary family $\Theta$ of dimplications $A \rightsquigarrow B$ {\it based} on $Q$, i.e. all occuring premises $A$ and conclusions $B$ are subsets of $Q$. Defining

(6) \quad ${\cal K}(\Theta) : = \{S \subseteq Q| \ \forall (A \rightsquigarrow B) \in \Theta : A \cap S = \emptyset \Ra B \cap S = \emptyset \}$

as the set of all $\Theta${\it -closed} subsets of $Q$ it is easy to see and well-known (Section 5) that ${\cal K}(\Theta)$ is a Knowledge Space. Given any Knowledge Space ${\cal K}$, there always are (many) families $\Theta$ such that ${\cal K} = {\cal K}(\Theta)$. In this case one calls $\Theta$ a {\it dimplication base} of ${\cal K}$. Two families of dimplications $\Theta_1$ and $\Theta_2$ are {\it equivalent} if ${\cal K}(\Theta_1) = {\cal K}(\Theta_2)$. Let us convince ourselves that for each dimplication base $\Theta$ of ${\cal K}$ there is a subset $\Theta' \subseteq PrimeDimp({\cal K})$ which is equivalent to $\Theta$. This will imply that $\Theta'$, and a fortiori $PrimeDimp({\cal K})$, is a dimplication base of ${\cal K}$. Indeed, any dimplication $A \rightsquigarrow \{b_1, \cdots, b_s\}$ of $\Theta$ is equivalent to the set of dimplications $A \rightsquigarrow b_1, \cdots, A \rightsquigarrow b_s$, which in turn is equivalent to $A_1 \rightsquigarrow b_1, \cdots, A_s \rightsquigarrow b_s$ where $A_i \subseteq A$ is any minimal subset such that $A_i \rightsquigarrow b_i$ still holds in ${\cal K}$. In other words, $A_i \rightsquigarrow b_i$ is a prime dimplication. Thus let $\Theta'$ be the collection of all arising prime dimplications $A_i \rightsquigarrow b_i$ as $A \rightsquigarrow \{b_1, \cdots, b_s\}$ ranges over $\Theta$.

As will be seen in Section 7, even large sets PrimeDimp$({\cal K})$ can be calculated fast based on Theorem 1. The bottleneck is rather calculating an equivalent {\it small} base of dimplications (which benefits the compression algorithms to be discussed in Sections 4 and 6).

{\bf 3.5} A Knowledge Space $(Q, {\cal K})$ that satisfies (A) from Section 2 is called a {\it Learning  Space}. It is well-known that Learning Spaces are exactly the knowledge structures satisfying (A) and (LC). In particular let $(Q, {\cal K})$ be a Learning Space and let $P$ be an atom. By (A) there is $m \in P$ such that $P \setminus \{m\} \in {\cal K}$. If there was $m'\neq m$ such that also $P \setminus \{m'\} \in {\cal K}$ then $(P \setminus \{m\}) \cup (P \setminus \{m'\}) = P$ contradicts $P$'s irreducibility. This unique $m$ is called the {\it color} of $P$. Using again the irreducibility of $P$ one sees that $P$ is an atom at $m$, and only at $m$. In other words, each Learning Space satisfies\footnote{This, and the converse implication, was first shown in Koppen [1993].} 

(LS) \quad Atoms$(m) \cap \mbox{Atoms}(m') = \emptyset$ for all $m \neq m'$.

This e.g. takes place for $(Q_2, {\cal K}_2)$ in Section 2; in fact one reads from Figure 2 that Atoms$(a) =\{ae\}$, Atoms$(b) = \{abe\}$, Atoms$(c) = \{cd, abce\}$, Atoms$(d) = \{d\}$, Atoms$(e) = \{e\}$. (In contrast, in ${\cal K}_1$ of Section 3 we had Atoms$(c) \cap \mbox{Atoms}(d) = \{cd\} \neq \emptyset$.) Apart from the prime dimplications of ${\cal K}_2$ used in Section 2, there are other ones; using Theorem 1 one calculates that PrimeDimp$({\cal K}_2)$ consists\footnote{This matches the list in [FD, p.299] where instead of (say) $bd \rightsquigarrow c$ the notation $\{b, d\} {\cal P}c$ is used.} of
$$e \rightsquigarrow a, \ \ a \rightsquigarrow b, \ \ bd \rightsquigarrow c, \ \ e \rightsquigarrow b, \ \ ad \rightsquigarrow c, \ \ de \rightsquigarrow c.$$

{\bf 3.6} We mention that in Combinatorics and Operations Research Learning Spaces are called {\it antimatroids}. A prime dimplication $A \rightsquigarrow b$ of a Learning Space would be referred to as {\it rooted circuit} $(A \cup \{b\}, b)$. As opposed to general Knowledge Spaces, for Learning Spaces ${\cal K}$ the set PrimeDimp$({\cal K})$ has a telltale structure. In terms of antimatroids it reads as follows. Let ${\cal R}{\cal C}$ be any family of ``rooted sets'' $(C, r)$, i.e. each $C$ is a nonempty subset of some domain $Q$ and $r \in C$. Then ${\cal R}{\cal C}$ is the family of all rooted circuits\footnote{As opposed to [YHM], the so called ``critical circuits'' [KLS, p.31] do not enter our framework. They have to do with linear orderings of $Q$. Details can be found in [W2].} of some antimatroid ${\cal K} \subseteq {\cal P}(Q)$ iff these conditions hold [Di]:

(7a) \quad $(C_1, r), (C_2, r) \in {\cal R}{\cal C}$ and $C_1 \subseteq C_2$ implies $C_1 = C_2$

(7b) \quad for $(C_1, r_1), (C_2, r_2) \in {\cal R} {\cal C}$ with $r_1 \in C_1 \setminus \{r_2\}$ there exists a\\
\hspace*{1cm} $(C_3, r_2)$ with $C_3 \subseteq (C_1 \cup  C_2) \setminus \{r_1\}$.

It is an exercise to verify that the six rooted circuits matching the prime dimplications above satisfy (7a) and (7b). 


\section{Compression of Knowledge Spaces using more\\
 subtle wildcards}

We proceed similarly to Section 2, i.e. queried dimplications that are accepted by the expert need to be imposed on a growing number of $012$-rows. The main novelty (4.1) will be the introduction of a more sophisticated wildcard besides the still useful don't care symbol ``2''. Correspondingly our $012$-rows get generalized to $012e$-rows. As to 4.2 and 4.3, see the Introduction.

{\bf 4.1} As often, a nontrivial toy example is worth more than excessive terminology. Thus we take $Q = \{1, 2, \ldots, 10\}$ as domain on which we shall construct a Knowledge Space ${\cal K}_3 \subseteq {\cal P}(Q)$. As in Section 2 we start with ${\cal P}(Q) = (2, 2, \cdots, 2)$. Suppose the dimplication $\{2, 3, 6\} \rightsquigarrow \{4, 7\}$, or briefly $236 \rightsquigarrow 47$, is to be imposed on $(2, 2, \cdots, 2)$. Clearly, all $S \in r_1$ in Table 4 below satisfy $236 \rightsquigarrow 47$. The other sets $S$ satisfying $236 \rightsquigarrow 47$ are exactly the ones in
$${\cal F}: = \{S \in {\cal P}(Q): S \cap \{2, 3, 6\} \neq \emptyset \}.$$
One could represent ${\cal F}$ as union of these three $012$-rows:

\begin{tabular}{|c|c|c|c|c|c|c|c|c|c|} \hline
$2$ & ${\bf 1}$ & 2 & 2 & 2 & 2 & 2 & 2& 2 & 2 \\ \hline
$2$ & $2$ & ${\bf 1}$ & 2 & 2 & 2 & 2 & 2& 2 & 2 \\ \hline
$2$ & $2$ & 2 & 2 & 2 & ${\bf 1}$ & 2 & 2 & 2 & 2 \\ \hline
\end{tabular}

Trouble is, these rows are not mutually disjoint; for instance $\{2, 3, 6\}$ even belongs to all three of them. Disjointness being essential we rather write ${\cal F} = r_2 \uplus r_3 \uplus r_4$ as shown in Table 3. The boldface pattern on $r_2, r_3, r_4$ has been coined, by obvious reasons, the {\it Flag of Papua} in other publications of the author. (This is just a handy name for the visualization of a well-known propositional tautology.) The Flag of Papua will reoccur later in other guise, but at the present stage we discard it by simply substituting\footnote{We switched the order of $r_1$ and $r_5$ by pedagogical reasons; i.e. we wish to continue with $r_5$ right away.} row $r_5$ for $r_2, r_3, r_4$. The $e$-wildcard $(e, e, e)$ by definition means that in  this area every bitstring contained in $r_5$ must have at least one 1 (and thus $r_5  = {\cal F}$). If several $e$-wildcards occupy the same {\it $012e$-row}, they get distinguished by subscripts.
For instance, let the next dimplication to be imposed be $\{5, 10\} \rightsquigarrow \{4\}$. This replaces $r_5$ by the disjoint union of $r_6 : = \{S \in r_5: S \cap \{5, 10\} \neq \emptyset \}$ and $r_7: = \{S \in r_5: S \cap \{5, 10, 4\} = \emptyset \}$. The subscripts of $e_1 e_1 e_1$ can be dropped in $r_7$ since it contains only one $e$-bubble.

Suppose the third dimplication to be imposed on the top row $r_6$ is $\{6, 9, 10\} \rightsquigarrow \{7\}$. That will be more cumbersome because $\{6, 9, 10\}$ clashes with both $(e_1, e_1, e_1)$ and $(e_2, e_2)$! For starters we decompose $r_6$ as $r_6 = {\cal F}_1 \cup {\cal F}_2$ where
$$\begin{array}{lll}
{\cal F}_1 & : = & \{S \in r_6 : S \cap \{6, 9, 10\} = \emptyset \}, \\
\\
{\cal F}_2 & : = & \{S \in r_6 : S \cap \{6, 9, 10\} \neq \emptyset \}. \end{array}$$
If one forces the second component of $(e_2, e_2)$ to $0$ then the wildcard becomes $(1,0)$. Similarly, forcing the last component $(e_1, e_1, e_1)$ to 0 yields $(e, e, 0)$. It is now clear that all $S \in {\cal F}_1$ satisfying $\{6, 9, 10\}\rightsquigarrow \{7\}$ are comprised in the $012e$-row $r_8$.  In order to impose $\{6, 9, 10\} \rightsquigarrow \{7\}$ upon ${\cal F}_2$ we first put ${\cal F}_2 = {\cal F}'_2 \cup {\cal F}''_2$ where
$$\begin{array}{lll}
{\cal F}'_2 & : = & \{S \in {\cal F}_2: 9 \in S\}, \\
\\
{\cal F}'' _2 & : = & \{S \in{\cal F}_2 : 9 \not\in S\}. \end{array}$$
Evidently all sets in $r_9 : = {\cal F}'_2$ satisfy $\{6, 9, 10\} \rightsquigarrow \{7\}$. As to ${\cal F}_2''$, it can be rewritten as ${\cal F}''_2 = \{S \in r_6 : 9 \not\in S$ and $S \cap \{6, 10\} \neq \emptyset \}$. If only $\{6, 10\}$ alone was concerned  we could represent the nonempty subsets of $\{6, 10\}$  by the Flag of Papua:

\begin{center}
\begin{tabular}{|c|c|} \hline
6 & 10 \\ \hline \hline
${\bf 1}$ & ${\bf 2}$ \\ \hline
${\bf 0}$ & ${\bf 1}$ \\ \hline \end{tabular}

Table 2
\end{center}

Unfortunately the component of $r_6$ indexed by 6 is the last $e_1$ of ($e_1, e_1, e_1)$. If it is forced to 1 (in view of the top left entry of the Flag of Papua) the wildcard turns to $(2,2,1)$. Likewise the component of $r_6$ indexed by 10 is the last $e_2$ in $(e_2,e_2)$. ``Forcing'' it to 2 (in view of the top right entry of the Flag of Papua) is no restriction at all; so $(e_2,e_2)$ simply remains $(e_2, e_2) = (e, e)$. It is now clear that the set system $\{S \in {\cal F}_2'' : 6 \in S\}$ coincides with row $r_{10}$. As to the second row $(0,1)$ in Table 2, forcing 0 upon the last entry of $(e_1, e_1, e_1)$ yields $(e, e, 0)$, and forcing 1 upon the last entry of $(e_2, e_2)$ yields $(2,1)$.

\begin{tabular}{l|c|c|c|c|c|c|c|c|c|c|l}
& 1 & 2 & 3 & 4 & 5 & 6 & 7 & 8 & 9 & 10 & \\ \hline 
&   &   &   &   &   &   &   &   &   &   & \\ \hline
$r_1 =$ & 2 & 0 & 0 & 0 & 2 & 0 & 0 & 2 & 2 & 2& \\ \hline 
$r_2 =$ & 2 & ${\bf 1}$ & ${\bf 2}$ & 2 & 2 & ${\bf 2}$ & 2 & 2 & 2 & 2 & \\ \hline
$r_3 =$ & 2 & ${\bf 0}$ & ${\bf 1}$ & 2 & 2 & ${\bf 2}$ & 2 & 2 & 2 & 2 & \\ \hline
$r_4 =$ & 2 & ${\bf 0}$ & ${\bf 0}$ & 2 & 2 & ${\bf 1}$ & 2 & 2 & 2 & 2 & \\ \hline
&   &   &   &   &   &   &   &   &   &   & \\ \hline
$r_5=$ & 2 & $e$ & $e$ & 2 & 2 & $e$ & 2 & 2 & 2& 2 & pending dimp. 2\\ \hline
$r_1=$ & 2 & 0 & 0 & 0 & 2 & 0& 0 & 2 & 2 & 2& pending dimp. 2\\ \hline 
&   &   &   &   &   &   &   &   &   &   & \\ \hline
$r_6=$ & 2 & $e_1$ & $e_1$ & 2 & ${\bf e}_2$ & $e_1$ & 2& 2 & 2 & ${\bf e_2}$ & pending dimp. 3\\ \hline
$r_7=$ & 2 & $e$ & $e$ & 0 & ${\bf 0}$ & $e$ & 2& 2 & 2 & ${\bf 0}$ & pending dimp. 3\\ \hline
$r_1=$ & 2 & $0$ & $0$ & 0 & $2$ & $0$ & 0 & 2 & 2 & $2$ & pending dimp. 2\\ \hline
&   &   &   &   &   &   &   &   &   &   & \\ \hline
$r_8=$ & 2 & $e$ & $e$ & 2 & 1 & 0 & 0& 2 & 0 & 0 & \\ \hline
$r_9=$ & 2  & $e_1$ & $e_1$ & 2 & $e_2$ & $e_1$ & 2 & 2 & ${\bf 1}$ & $e_2$ & \\ \hline
$r_{10}=$ & 2 & $2$ & $2$ & 2 & $e$ & ${\bf 1}$ & 2 & 2 & 0 & ${\bf e}$ & \\ \hline
$r_{11}=$ & 2 & $e$ & $e$ & 2 & $2$ & ${\bf 0}$ & 2 & 2 & 0 & ${\bf 1}$ & \\ \hline
$r_7=$ & 2 & $e$ & $e$ & 0 & $0$ & $e$ & 2 & 2 & 2 & $0$ & pending dimp. 3 \\ \hline
$r_1=$ & 2 & 0 & 0 & 0 & 2 & 0 & 0 & 2 & 2 & 2 & pending dimp. 2 \\ \hline
\end{tabular}

Table 3: Starting to compress ${\cal K}_3$ with $012e$-rows

This explains why the set system $\{S \in {\cal F}_2'': 6 \not\in S\}$ coincides with row $r_{11}$. The boldface entries in $r_{10}, r_{11}$ are a reflection of the Flag of Papua in Table 3. The pending dimplication in $r_8$ to $r_{11}$ is the fourth one.

\includegraphics{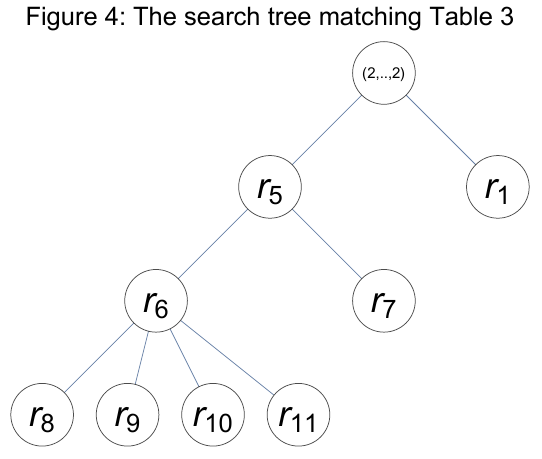}

Suppose altogether these 12 dimplications need to be imposed:

\hspace*{.9cm} $\{2, 3, 6\} \rightsquigarrow \{4, 7\}, \  \ \{5, 10\} \rightsquigarrow \{4\}, \ \ \{6, 9, 10\} \rightsquigarrow \{7\}, \ \ \{3\} \rightsquigarrow \{4\}, \ \ \{5, 7\} \rightsquigarrow \{4\}$,

\hspace*{.9cm} $\{7,8\} \rightsquigarrow \{4\}, \ \ \{8, 10\} \rightsquigarrow \{4\}, \ \ \{1\} \rightsquigarrow \{6\}, \ \  \{1, 3\} \rightsquigarrow \{4, 6, 7\}$, \ \ 

\hspace*{.9cm} $\{1, 10\} \rightsquigarrow \{6, 7\}, \ \ \{2,6, 10\} \rightsquigarrow \{7\}, \ \ \{3, 6, 9\} \rightsquigarrow \{4, 7\}$

In ways similar\footnote{All possible cases arising when imposing a dimplication onto an arbitrary $012e$-row have been dealt with in a dual framework in [W1]. As to ``dual'', see Section 5.} to the above one gets the twenty three $012e$-rows in Table 4. Summing the cardinalities of the rows in Table 4 gives

(8) \quad $|{\cal K}_3| = 2^5 + 2^4 + \cdots + 2^2 + 2^4 (2^2-1) = 377$.

All but the last row in Table 4 happen to be $012$-rows (which still beats one-by-one enumeration). One can easily come up with instances having more ```proper'' $012e$-rows. See also Section 7.

\begin{tabular}{|c|c|c|c|c|c|c|c|c|c|}
1 & 2 & 3& 4 & 5 & 6 & 7 & 8 & 9 & 10\\ \hline
&   &   &   &   &   &   &   &   & \\ \hline
0 & 2 & 0 & 0 & 2 & 0 & 0 & 2 & 2 & 2 \\ \hline
1 & 0 & 0 & 0 & 2 & 0 & 0 & 2 & 2 & 2  \\ \hline
1 & 1 & 0 & 0 & 2 & 0 & 0 & 2 & 0 & 2 \\ \hline
1 & 1 & 0 & 0 & 2 & 0 & {\bf 2} & 2 & 1 & 2 \\ \hline
1 & 2 & 0 & 0 & 2 & 1 & {\bf 2} & 2 & 2 & 2\\ \hline
0 & 2 & 1 & 0 & 0 & 0 & 0 & 2 & 2 & 2\\ \hline
1 & 2 &1 & 0 & 0 &2 & 0 & 2 & 2 & 2\\ \hline
1 & 1 &  1 & 0 & 0 & 0 & {\bf 1} & 2 & 1 & 0 \\ \hline
1 & 2 & 1 & 0 & 0 & 1 & {\bf 1} & 2 & 2 &0 \\ \hline
0 & 2 & 1 & 2 & 0 & 0 & {\bf 1} & 2 & 2 & 1 \\ \hline
1 & 2 & 1 & 2 & 0 & 2 & {\bf 1} & 2 & 2 & 1\\ \hline 
0 & 2 & 1 & 0 & 1 & 0 & 0 & 0 & 2 & 2 \\ \hline
1 & 2 & 1 & 0 & 1 & 2 & 0 & 0 & 2 & 2 \\ \hline
1 & 1 & 1 & 0 & 1 & 0 & {\bf 1} & 0 & 1 & 0 \\ \hline
1 & 2 & 1 & 0 & 1 & 1 & {\bf 1} & 0 & 2 & 0 \\ \hline
0 & 2 & 1 & 2 & 1 & 0 & {\bf 1} & 0  & 2 & 1 \\ \hline
1 & 2 & 1 & 2 & 1 & 2 & {\bf 1} & 0 & 2 & 1 \\ \hline
0 & 2 & 1 & 2 & 1 & 0 & 0 & 1 & 2 & 0 \\ \hline
0 & 2 & 1 & 2 & 1 & 0 & {\bf 2} & 1 & 2 & 1 \\ \hline
1 & 0 & 1 & 2 & 1 & 0 & 0 & 1 & 2 & 0 \\ \hline
1 & 1 & 1 & 2 & 1 & 0 & 0 & 1 & 0 & 0 \\ \hline
1 & 1 & 1  & 2 & 1 & 0 & {\bf 2} & 1 & 1 & 0\\ \hline
1 & 2 & 1 & 2 & 1 & $e$ & {\bf 2} & 1 & 2 & $e$\\ \hline
\end{tabular}

Table 4: Final compression of the Learning Space ${\cal K}_3$ with $012e$-rows

{\bf 4.2} Up to the duality discussed in Section 5, it is shown in [W1, Thm.2] that for a given family $\Theta$ of $h$ dimplications based on $[w]: = \{1, 2, \cdots, w\}$ the Knowledge Space ${\cal K}(\Theta)$ can be represented as a disjoint union of $R$ many $012e$-rows in time $O(Rh^2 w^2)$. Only $R \leq N : = |{\cal K}(\Theta)|$ can be guaranteed but in practise often $R \ll N$, as will be seen in Section 7.

A compression of a Knowledge Space ${\cal K}$ as in Table 4 allows all kinds of statistical analysis which would be cumbersome if only ${\cal B}({\cal K})$ was known. For instance, what is the probability $p$ that a student managing tasks 3 and 4 fails both 9 an 10? Assuming\footnote{This is an oversimplication, aimed at keeping the calculations elementary.} that for a random student any knowledge state is equiprobable, the question can be answered by first listing all knowledge states containing 3 and 4. Viewing rows 10, 11 and 16 to 23 in Table 4 it is clear that the latter set of knowledge states is represented by the left part of Table 5. They number to 77.

\begin{tabular}{|c|c|c|c|c|c|c|c|c|c|c|c|c|c|c|c|c|c|c|c|c|}
1 &  2& 3 & 4 & 5 & 6  & 7 & 8 & 9 & 10 & & 1 & 2& 3 & 4 & 5 & 6 & 7 & 8 & 9 & 10\\ \hline
  &   &   &   &   &    &   &   &   &   & &   &   &  &   &    &   &   &   &   & \\ \hline
0 & 2 &  ${\bf 1}$ & ${\bf 1}$ & 0 & 0 & 1 & 2 & 2 & 1 & &    &  &   &  &    &    &   &   &   & \\ \hline
1 & 2 & ${\bf 1}$ & ${\bf 1}$ & 0 & 2 & 1 & 2 & 2 & 1 & &   &  &   &  &    &    &   &   &   & \\ \hline
0 & 2 & ${\bf 1}$ & ${\bf 1}$ & 1 & 0 & 1 & 0 & 2 & 1 & &   &  &   &  &    &    &   &   &   & \\ \hline
1 & 2 & ${\bf 1}$ & ${\bf 1}$ & 1 & 2 & 1 & 0 & 2 & 1 & &   &  &   &  &    &    &   &   &   & \\ \hline
0 & 2 & ${\bf 1}$ & ${\bf 1}$ & 1 & 0 & 0 & 1 & 2 & 0 & &  0 & 2 & ${\bf 1}$ & ${\bf 1}$ & 1 & 0 & 0 & 1 & ${\bf 0}$ & ${\bf 0}$ \\ \hline
0 & 2 & ${\bf 1}$ & ${\bf 1}$ & 1 & 0 & 2 & 1 & 2 & 1 & &    &   &   &   &   &   &   &   &    & \\ \hline
1 & 0 & ${\bf 1}$ & ${\bf 1}$ & 1 & 0 & 0 & 1 & 2 & 0 & & 1 & 0 & ${\bf 1}$ & ${\bf 1}$ & 1 & 0 & 0 & 1 & ${\bf 0}$ & ${\bf 0}$ \\ \hline
1 & 1 & ${\bf 1}$ & ${\bf 1}$ & 1 & 0 & 0 & 1 & 0 & 0 & & 1 & 1 & ${\bf 1}$ & ${\bf 1}$ & 1 & 0 & 0 & 1 & ${\bf 0}$ & ${\bf 0}$ \\ \hline
1 & 1 & ${\bf 1}$ & ${\bf 1}$ & 1 & 0 & 2 & 1 & 1 & 0 & &   &   &   &   &   &   &   &   &   & \\ \hline
1 & 2 & ${\bf 1}$ & ${\bf 1}$ & 1 & $e$ & 2 & 1 & 2 & $e$ & & 1 & 2 & ${\bf 1}$ & ${\bf 1}$ & 1 & 1 & 2 & 1 & ${\bf 0}$ & ${\bf 0}$ \\ \hline
  \end{tabular}

Table  5: Statistical analysis of a Knowledge Space

The knowledge states containing 3 and 4 but avoiding 9 and 10 are enshrined in the right part of Table 5. They number to 8, and so $p = 8/77$.

{\bf 4.3} From Table 4 we can also calculate Atoms$(7)$ (say) as follows. Since $7 \in X$ for all $X \in \, \mbox{Atoms}(7)$ we see that Atoms$(7)$ is contained in the union of  the {\it candidate rows} in Table 4, i.e. whose seventh entry is $1$ or $2$ (rendered boldface). In fact, if RowMin denotes the set of all {\it row-minimal} sets of candidate rows, then by definition of Atoms$(7)$ we have Atoms$(7) \subseteq \ \mbox{RowMin}$. We claim that the members of RowMin are the bitstrings in Table 6:

\begin{tabular}{l|c|c|c|c|c|c|c|c|c|c|}
 & 1 & 2 & 3& 4 & 5 & 6 & 7 & 8 & 9 & 10 \\ \hline
 &  &   &   &  &   &   &   &   &  & \\ \hline
 $b=$ & 1 & 1 & 0 & 0 & 0 & 0 & 1 & 0 & 1& 0 \\ \hline
 $e=$ & 1 & 0 & 0 & 0 & 0 & 1 & 1 & 0 & 0 & 0 \\ \hline
  & 1 & 1 & 1 & 0 & 0 & 0 & 1 & 0 & 1 & 0 \\ \hline
  & 1 & 0 & 1 & 0 &  0 & 1 & 1 & 0 & 0 & 0 \\ \hline 
  $c=$ & 0 & 0 & 1 & 0 & 0 & 0 & 1 & 0 & 0 & 1 \\ \hline 
  & 1 & 0 & 1 & 0 & 0 & 0 & 1 & 0 & 0 & 1\\ \hline
  & 1 & 1 & 1 & 0 & 1 & 0 & 1 & 0 & 1 & 0 \\ \hline
  & 1 & 0 & 1 & 0 & 1 & 1 & 1 & 0 & 0 & 0 \\ \hline 
  & 0 & 0 & 1 & 0 & 1 & 0 & 1 & 0 & 0 & 1\\ \hline
  & 1 & 0 & 1 & 0 & 1 & 0 & 1 & 0 & 0 & 1\\ \hline
  & 0 & 0 & 1 & 0 & 1 & 0 & 1 & 1 & 0 & 1\\ \hline
  & 1 & 1 & 1 & 0 & 1 & 0 & 1 & 1 & 1 & 0\\ \hline
  & 1 & 0 & 1 & 0 & 1 & {\bf 1} & 1 & 1 & 0 & {\bf 0} \\ \hline
  & 1 & 0 & 1 & 0 & 1 & {\bf 0} & 1 & 1 & 0 & {\bf 1} \\ \hline \end{tabular}

Table 6: The row-minimal sets of the candidate rows in Table 4

Indeed, each candidate row without $e$-symbol has exactly one minimal member. It is obtained by turning 2 at position 7 to 1 (or leaving it 1 if it is $1$) and turning all other 2's to 0. The last candidate row features $(e,e)$. It is clear why it has {\it two} minimal members, i.e. the last two rows in Table 6. (Generally, a candidate row with $t$ many $e$-wildcards of length $\varepsilon_1, \cdots, \varepsilon_t$ respectively has $\varepsilon_1 \varepsilon_2 \cdots \varepsilon_t$ many minimal members.) By sieving the (absolutely) inclusion-minimal sets among the row-minimal sets yields Atoms$(7) = \{b, e, c\}$. In the same way one determines Atoms$(m)$ for $m \in Q\setminus \{7\}$. The result is given in Figure 5. The boldface number in each set ($=$ join irreducible) $P$ points out the color of $P$.

\begin{center}
\includegraphics[scale=0.6]{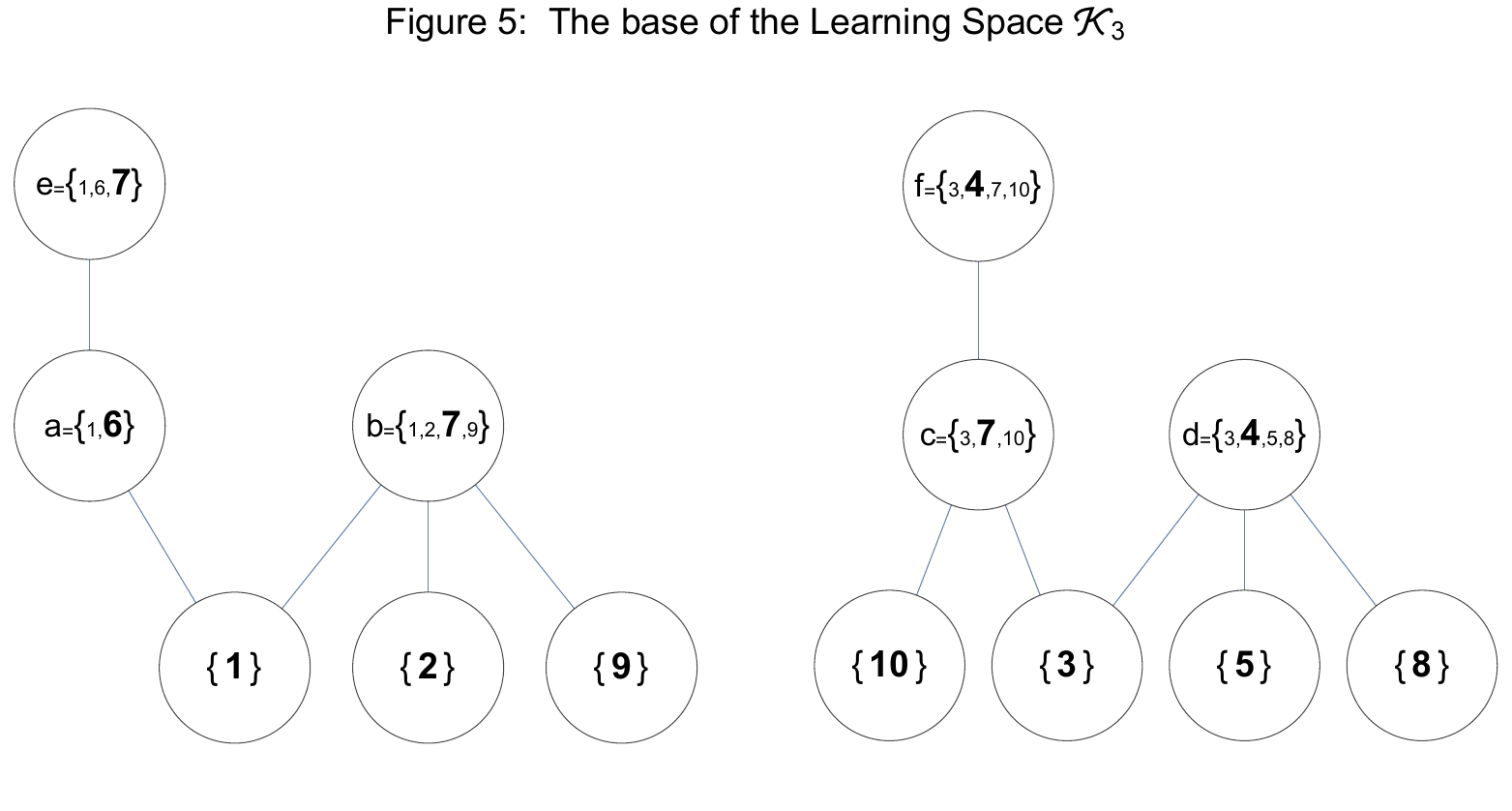}
\end{center}

\section{From Knowledge Spaces to closure spaces to lattices}

In order to present another type of compression in Section 6 we need some more theory.
A {\it closure space} (or closure system [G, p.47]) with domain $Q$ is a pair $(Q, {\cal C})$ such that ${\cal C} \subseteq {\cal P}(Q)$ contains $\emptyset$ and $Q$, and such that

(9) \quad $X \cap Y \in {\cal C}$ \ for all \ $X, Y \in {\cal C}$.

The sets in ${\cal C}$ are called {\it closed} sets. As is well-known, coupled to ${\cal C}$ is a {\it closure operator} which maps any set $S \subseteq Q$ to the smallest closed superset of $S$, i.e. to $\ol{S} : = \bigcap \{X \in {\cal C}: S \subseteq X\}$.
For subsets $A, B$ of a set $Q$ the expression $A \ra B$ is called an {\it implication}. It {\it holds} in a subset $X \subseteq Q$ if

(10) \quad $A \subseteq X \ \Ra \ B \subseteq X$. \ (Equivalently: $A \not\subseteq X$ or $B \subseteq X$.)

An implication, or more generally a conjunction of implications is nothing else than a {\it pure Horn formula}, and thus belongs to the domain of Boolean logic [CH]. However, it often pays to have a smoother terminology and symbolism. That point of view originated in Formal Concept Analysis (FCA) some thirty years ago (see [GO]) and is also adopted in [W2]. By definition $A \ra B$ holds in $(Q, {\cal C})$ if $A \ra B$ holds for all $X \in {\cal C}$. Conversely, if one starts with any family $\Sigma$ of implications based on a set $Q$ and defines
$${\cal C}(\Sigma): = \{ X \subseteq Q | \ \forall (A \ra B) \in \Sigma: \ A \subseteq X \ \Ra \ B \subseteq X \},$$
then ${\cal C}(\Sigma)$ is a closure system. Given any Closure Space $(Q,{\cal C})$, a family $\Sigma$ with ${\cal C} = {\cal C}(\Sigma)$ is called an {\it implication base} of ${\cal C}$. 
Two families $\Sigma$ and $\Sigma'$ are {\it equivalent} if ${\cal C}(\Sigma) = {\cal C}(\Sigma')$. Among all implicational bases of ${\cal C}$ there is a unique ``canonical'' implicational base $\Sigma_{GD}$ (Guiges-Duquenne) which has minimum cardinality and other extra features [GO, chapter 3].

There clearly is a link to dimplications. More specifically, it is based on the equivalence 

(11) \quad $A \rightsquigarrow B$ holds in $S \ \ \Leftrightarrow \ \ A \ra B$ holds in $Q \setminus S$,

which we leave as an exercise for the reader, and which entails that every statement true in one framework can be dualized\footnote{In doing so so the statement may however become clumsy. A case in point is Theorem 1 whose dualization [W2, Thm.4] has an ``uglier'' proof. Another case in point is Dowling's algorithm (Section 3) which is also crisper in a Knowledge Space setting. On the other hand, one may argue that the very concept of ``implication'' is more intuitive than the concept of ``dimplication''.} to the other one. We point out however that historically more research has been done in the framework of closure spaces.

{\bf 5.1} We refer the reader to e.g. [G] for basic facts on {\it lattices}. In particular we shall write $JI({\cal L})$ and $MI({\cal L})$ respectively for the sets of {\it join} and {\it meet-irreducibles} of a lattice ${\cal L}$. As is well-known, each Knowledge Space $(Q, {\cal K})$, viewed as set system $({\cal K}, \subseteq)$ partially ordered by inclusion, provides an example of a lattice. The joins and meets are given by

(12) \quad $K \vee L = K \cup L$ and $K \wedge L = (K \cap L)^\circ$ \ \ (see (4)).

Dually, for each closure space $(Q, {\cal C})$ the partially ordered set system $({\cal C}, \subseteq)$ is a lattice with meets and joins given by

(13) \quad $X \wedge Y = X \cap Y \ \mbox{and} \ X \vee Y = \ol{X \cup Y}.$

This kind of lattice is not special; {\it each} lattice ${\cal L}$ is isomorphic to one of kind (13) (or (12)). Namely, putting $JI(X): = \{P \in JI({\cal L}): P  \leq X\}$ $(X \in {\cal L})$ the set system

(14) \quad ${\cal C}({\cal L}) : = \{JI(X): X \in {\cal L}\}$

turns out to be a closure system on $JI({\cal L})$ which, viewed as a lattice, is isomorphic to ${\cal L}$. Thus a family $\Sigma$ of implications $A \ra B$ $(A, B \subseteq JI({\cal L}))$ is called an {\it implication base of a lattice} ${\cal L}$ if ${\cal C}(\Sigma) = {\cal C}({\cal L})$. For many types of lattices ${\cal L}$ a lot about the implication bases of ${\cal L}$ are known [W2]. In 5.2 we focus on a particular type of lattice that in Section 6 will be related to Learning Spaces.

{\bf 5.2} For any lattice ${\cal L}$ each $P \in JI({\cal L})$ has a unique lower cover $P_\ast$ and each $M \in MI({\cal L})$ has a unique upper cover $M^\ast$. The following relations between join and meet irreducibles are essential\footnote{The particular symbols $\uparrow, \downarrow$, and $\updownarrow$ are adopted from FCA, although they don't appear in [GO] anymore.} in the structure theory of finite lattices.

(15) \quad $P \uparrow M \ : \Leftrightarrow \ P \vee M = M^\ast$

\hspace*{1cm} $P \downarrow M \ : \Leftrightarrow \ P \wedge M = P_\ast$

\hspace*{1cm} $P \updownarrow M \ : \Leftrightarrow \ P \uparrow M$ and $P \downarrow M$

As is well-known and easily seen, for $P \in JI({\cal L})$ fixed, the elements $M \in MI({\cal L})$ with $P \updownarrow M$ are exactly the elements $X \in {\cal L}$ which are maximal with respect to the property that $X \geq P_\ast$ but $X \not\geq P$. That leads us to the definition of {\it meet semidistributive} $(SD_\wedge)$ lattices:

(16) \quad A lattice ${\cal L}$ is $SD_\wedge$ iff for each $P \in JI({\cal L})$ there is a {\it unique} $M = M(P)$ with $P \updownarrow M$.

According to [JN] for each $SD_\wedge$ lattice ${\cal L}$ one can obtain a small implication base as follows. On $JI({\cal L})$ consider this binary relation: 

(17) \quad $P \mapsto R \ : \Leftrightarrow \ M(P) \wedge R = R_\ast$ \quad (i.e. $R \downarrow M(P)$)

The left of Figure 6 illustrates the relations between $P, P_\ast, M = M(P), M^\ast$ when $P \updownarrow M$ takes place. Furthermore it illustrates that $P \mapsto R$ entails more than $M \not\geq R$ (which is equivalent to $M \wedge R < R$), i.e. it entails $M \wedge R = R_\ast$.

\begin{center}
\includegraphics[scale=0.6]{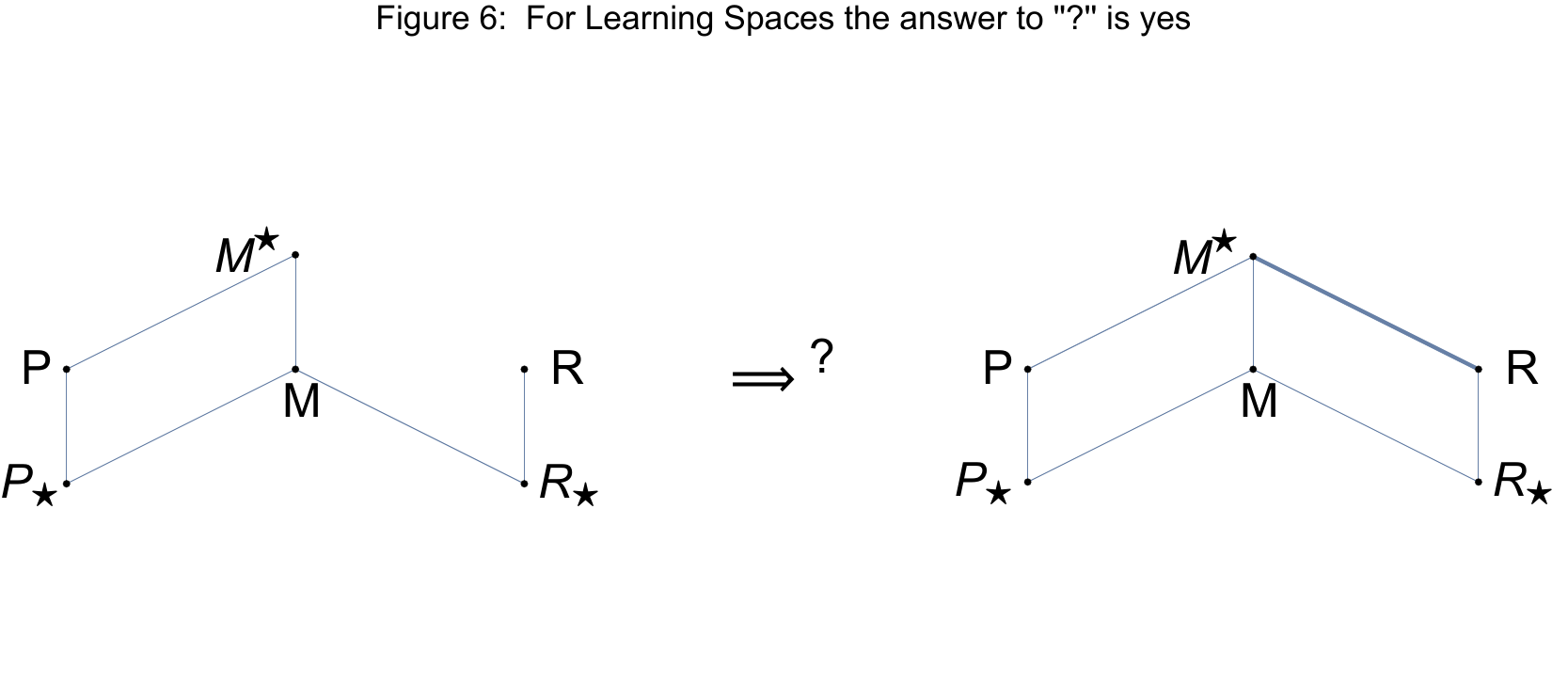}
\end{center}

It is natural to ask (see the ``?'' in Figure 6) under what circumstances the $R$ and $M^\ast$ on the left are forced to be related as $R \leq M^\ast$. The caption of Figure 6 forecasts our answer in Section 6.

For $P \in JI({\cal L})$ we write $lcov(P)$ for the set of lower covers of $P$ within the poset $(JI({\cal L}), \leq )$ (thus not within $({\cal L}, \leq)$). Here $\leq$ is the partial ordering of ${\cal L}$ restricted to the subset $JI({\cal L}) \subseteq {\cal L}$.  Observe that $lcov(P) = \emptyset$ if $P$ is a minimal member of $JI({\cal L})$. We put

(18) \quad $\Sigma_{po} : = \{\{P\} \ra lcov(P) : P \in JI({\cal L}), P \ \mbox{not minimal}\},$

where $po$ stands for implications forced by the mere {\it poset} structure of $JI({\cal L})$. Furthermore set

(19) \quad $\Sigma_{JN}: = \{\{P\} \cup lcov(R) \ra \{R\} : \  \ P, R \in JI({\cal L}) \ \mbox{and} \ P \mapsto R \}.$

By [JN, Thm.1] an implication base of the $SD_\wedge$-lattice ${\cal L}$ is given by $\Sigma ({\cal L}) : = \Sigma_{po} \cup \Sigma_{JN}$. If $|JI({\cal L})| = k$ then $|\Sigma({\cal L})| \leq k + k(k-1) =k^2$.  Let ${\cal D} {\cal G}({\cal L})$ be the directed graph with vertex set $JI ({\cal L})$ and an arc from $P$ to $R$ iff $P \mapsto R$ (as defined in (17)). The generally complicated structure of ${\cal D}{\cal G}({\cal L})$ drastically simplifies in the context of Section 6.

\section{Learning Spaces enjoy a second type of compression}

We are going to show that the theory of 5.2 nicely simplifies for particular $SD_\wedge$ lattices, i.e. lattices ${\cal L}$ that derive, in the sense of (12), from a Learning Space $(Q, {\cal K})$. To  begin with, let $P \in JI({\cal L})$ be arbitrary. Thus $P \in {\cal B} = {\cal B}({\cal K})$. Say $P$ has color $m$ (Section 3),  and so $P \setminus P_\ast = \{m\}$. By (3) the set $M: = \bigcup \{K \in {\cal K}: m \not\in K\}$ belongs to ${\cal L}$, and it evidently is the largest member of ${\cal L}$ that contains $P_\ast$ but not $m$. Hence $M = M(P)$ is the unique meet irreducible with $P \updownarrow M(P)$. (Notice $M(P)^\ast = M(P) \cup P = M(P) \cup \{m\}$ in view of (15).) This shows that ${\cal L}$ satisfies $SD_\wedge$. In fact, much more takes place:

(20) \quad In any lattice ${\cal L}$ derived from a Learning Space it follows\footnote{A priori more general, this follows in any {\it locally upper distributive} lattice ${\cal L}$. But any such lattice is ``isomorphic'' to a lattice derived from a Learning Space, see e.g. [KLS, p.77]. Thus we may as well give the proof in the current Learning Space setting.} from $P \mapsto R$ that $R \mapsto P$.

Pictorially the symmetry of the relation $\mapsto$ amounts to the validity of the implication at stake in Figure 6.

{\it Proof of (20)}. By definition $P \mapsto R$ means $M(P) \wedge R = R_\ast$. As before let $P = P_\ast \cup \{m\}$, and let $R = R_\ast \cup \{m_0\}$. We shall show that $m_0 =m$. Then $M(R) = \cup \{K \in {\cal K} : m_0 \not\in K \} = \cup \{K \in{\cal K} : m \not\in K\} = M(P)$, hence $M(R) \wedge P = M(P) \wedge P = P_\ast$, hence $R \mapsto P$.

As to showing $m_0 = m$, note that $m_0 \not\in M(P)$ since otherwise the inclusion $R_\ast \subseteq M(P)$ (which is implied by $M(P) \wedge R = R_\ast$) would entail $R \subseteq M(P)$ and whence the contradiction $M(P) \wedge R = R$. From $M(P) \in {\cal L}$ and $R_\ast \cup \{m_0\} \in {\cal L}$ follows $M(P) \cup \{m_0\} = M(P) \cup (R_\ast \cup \{m_0\}) \in {\cal L}$. Recall $m, m_0 \not\in M(P)$. If we had $m_0 \neq m$ then $M(P) \cup \{m_0 \} \neq M(P) \neq M(P) \cup \{m\}$ would imply $(M(P) \cup \{m_0\}) \cap (M(P) \cup \{m\}) = M(P)$, contradicting the meet irreducibility of $M(P)$. \quad $\square$

It follows from (20) that the directed graph ${\cal D}{\cal G}({\cal L})$  of 5.2 has its connected components corresponding to the meet-irreducibles, and each connected component is a (directed) clique. All cliques are singletons iff $\Sigma_{JN} = \emptyset$.  We mention in passing that this happens iff ${\cal L}$ is a distributive lattice.

{\bf Theorem 2:} Let $(Q,{\cal K})$ be the Learning Space determined by its explicitely given base ${\cal B} \subseteq {\cal P}(Q)$. If $|{\cal B}| =k$ then an at most $k^2$-element implication base $\Sigma ({\cal L})$ of ${\cal L} = {\cal K}$ (viewed as lattice as in (12)) can be calculated in time $O(k^3|Q|)$.

{\it Proof.} For each fixed $P \in {\cal B}$ determining the maximal members $P_0 \in {\cal B}$ contained in $P$ (i.e. its lower covers) costs $O(k^2|Q|)$. In view of (14) setting up $\Sigma_{po}$ hence costs $O(k^3|Q|$). The unique element of $P$ not in the union of $P$'s lower covers is the color of $P$. Any two base sets with the same color determine two implications of $\Sigma_{JN}$, see (19). It follows that the overall cost to calculate $\Sigma ({\cal L})$ remains $O(k^3|Q|)$. \hfill $\square$

{\bf 6.1} To fix ideas, take $E = [10]$ and consider  the Learning Space ${\cal L}={\cal K}_3$ generated by the 13 base sets in ${\cal B}$ displayed in Figure 5. Their colors are indicated in boldface. The only colors occuring more than once are 7 (thrice) and 4 (twice). Thus the nontrivial cliques of ${\cal D}{\cal G}({\cal L})$ have three and two elements respectively.

\begin{center}
\includegraphics{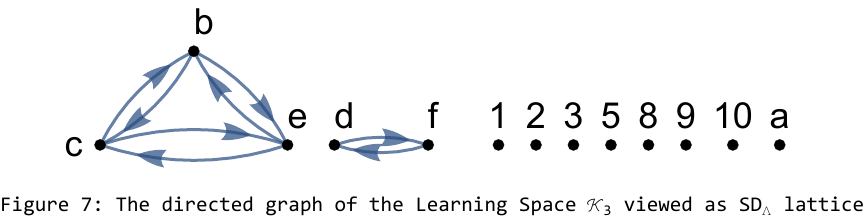}
\end{center}

Correspondingly there will be $2{3 \choose 2} + 2 {2 \choose 2}=8$ implications in $\Sigma_{JN}$:

$\begin{array}{ll}
\{ b, \{3\}, \{10\}\} \ra \{c\} & (\mbox{because of} \ b \mapsto c)\\
\\
\{c, \{1\}, \{2\}, \{9\}\} \ra \{b\} & (\mbox{because of} \ c \mapsto b) \\
\\
\{b,a\} \ra \{e\} & (\mbox{because of} \ b \mapsto e)\\
\\
\{e, \{1\}, \{2\}, \{9\}\} \ra \{b\} & (\mbox{because of} \ e \mapsto b)\\
\\
\{c,a\} \ra \{e\} & (\mbox{because of} \ c \mapsto e)\\
\\
\{e, \{3\}, \{10\}\} \ra \{c\} & (\mbox{because of} \ e \mapsto c)\\
\\
\\
\{d,c\} \ra \{f\} & (\mbox{because of} \ d \mapsto f) \\
\\
\{f, \{3\},  \{5\}, \{8\}\} \ra \{d\} & (\mbox{because of} \ f \mapsto d) 
\end{array}$

Thus $\Sigma ({\cal L}) = \Sigma_{po} \cup \Sigma_{JN}$ where $\Sigma_{po} = \{\{a\} \ra lcov(a), \ldots, f \ra lcov(f)\}$. Spelling out say $\{b\} \ra lcov(b)$ gives $\{b\} \ra \{\{1\}, \{2\}, \{9\} \}$. 
Feeding $\Sigma ({\cal L})$ to the implication $n$-algorithm of [W1] represents ${\cal L}$ as a disjoint union of eighteen $012n$-valued rows as follows:

\begin{tabular}{c|c|c|c|c|c|c|c|c|c|c|c|c|c|c} 
& 1 & 2 & 3 & 5 & 8 & 9 & 10 & $a$ & $b$ & $c$ & $d$ & $e$ & $f$ & \\ \hline
&   &   &   &   &   &   &    &     &     &     &     &     &  & \\ \hline 
$r_1=$ & 1 & 1 & 1 & 1 & 1& 1 & 1 & 1 & 1  & 1 & 1 & 1 & 1 & \ 1\\ \hline 
$r_2=$ & 1 & $n$ & 1 & 1 & 1 & $n$ & 1 & 1 & 0 & 1 & 1 & 1& 1& \ 3 \\ \hline 
$r_3=$ & 1 & 1 & 1 & $n$ & $n$ & 1 & 1 & 1 & 1 & 1  &0 & 1 & $n$ & \ 7 \\ \hline 
$r_4=$ & 1 & $n_1$ & 1 & $n_2$ & $n_2$ & $n_1$ & 1 & 1 & 0 & 1 & 0 & 1 & $n_2$ & \ 21 \\ \hline 
$r_5=$ & 1 & 1 & 1 & 1 & 1 & 1 & 0 & 1 & 1 & 0 & 1 & 1 & 0 & \ 1\\ \hline 
$r_6=$ & 1 & $n$ & 1 & 1 & 1 & $n$ & 0 & 1 & 0 & 0 & 1 & 1& 0 & \ 3\\ \hline 
$r_7=$ & 1 & 2 & 1 & 1 & 1 & 2 & 2& 1 & 0 & 0 & 1 & 0 & 0 & \ 8 \\ \hline 
$r_8 =$ & 1 & 1 & $n$ & 2 & 2 & 1 & $n$ & 1 & 1 & 0 & 0 & 1 & 0 & \  12 \\ \hline
$r_9=$ & 1 & $n_1$ & $n_2$ & 2 & 2 & $n_1$ & $n_2$ & 1 & 0 & 0 & 0 & 1 & 0 & \ 36 \\ \hline
$r_{10}=$ & 1 & 2 & 2 & 2 & 2 & 2 & 2 & 1 &0 & 0 & 0 & 0 & 0 & \ 64 \\ \hline 
$r_{11}=$ & 1 & 1 & 1 & 1 & 1 & 1 & 1 & 0 & 1 & 1& 1& 0 & 1 & \ 1 \\ \hline
$r_{12}=$ & $n$ & $n$ & 1 & 1 & 1 & $n$ & 1 & 0 & 0 & 1& 1 & 0 & 1 & \ 7 \\ \hline
$r_{13}=$ & 1 & 1 & 1 & $n$ & $n$ & 1 & 1 & 0 & 1 & 1 & 0 & 0 & $n$ & \ 7 \\ \hline 
$r_{14}=$ & $n_1$ & $n_1$ & 1 & $n_2$ & $n_2$ & $n_1$ & 1&  0 & 0 & 1 & 0 & 0 & $n_2$ & \ 49\\ \hline
$r_{15}=$ & 1 & 1 & 1 & 1 & 1 & 1 & 0 & 0 & 1 & 0 & 1 & 0 & 0 & \ 1 \\ \hline
$r_{16}=$ & 2 & 2 & 1 & 1& 1 & 2 & 2& 0 & 0 & 0 &1 & 0 & 0 & \ 16\\ \hline
$r_{17}=$ & 1 & 1 & $n$ & 2 & 2 & 1& $n$ & 0 & 1 & 0 & 0 & 0 & 0 & \ 12\\ \hline
$r_{18}=$ & 2 & 2 & 2 & 2 & 2 & 2 & 2& 0 & 0 & 0 & 0 & 0 & 0 & \ 128 \\ \hline 
 \end{tabular}

Table 7: Compression of the Learning Space ${\cal K}_3$ with $012n$-rows

This $n$-algorithm is very much the dual of the $e$-algorithm discussed in Section 4 in that now $(n, n, \cdots, n)$ by definition means: all bitstrings are allowed except for $(1, 1, \cdots, 1)$. For instance taking $n_1n_1= 00$ and $n_2n_2n_2 = 110$ in $r_4$ gives us one of 21 bitstrings contained in $r_4$, namely
$$(1, 0, 1, 1, 1, 0, 1, 1, 0, 1, 0, 1, 0) = \{1, 3, 5, 8, 10, a, c, e\} = : X.$$
Thus $X$ is one of $1+3+ \cdots + 128 = 377$ many $\Sigma ({\cal L})$-closed subsets of $JI({\cal L})$, whence $|{\cal L}| = 377$, which matches (8). If we replace each join irreducible in $X$ by the elements of the base set it represents we get the knowledge state
$$\{1\} \cup \{3\} \cup \{5\} \cup \{8\} \cup \{10\} \cup \{1,6\} \cup \{3, 7, 10\} \cup \{1, 6, 7\} = \{1, 3, 5, 8, 10, 6, 7\}.$$
While counting the number of knowledge states is as smooth in Table 7 as in Table 4, we see that the kind of processing done in Tables 5 and 6 is more cumbersome for the $012n$ type of compression.

\section{Numerical experiments}

Our computer experiments are threefold. The calculation of the Knowledge Space ${\cal K}(\Theta)$ generated by a family $\Theta$ of dimplications  (see Section 4) will be assessed in 7.1. The two ways of calculating a Knowledge Space from its base ${\cal B}$ (using Theorem 1 respectively Dowling's algorithm) will be compared in 7.2. Finally for {\it Learning} Spaces the two ways of compression (Section 4 or 6) will be pit against each other in 7.3.

{\bf 7.1} For various choices of parameters $w, h, a, b$ we randomly generated families $\Theta$ of $h$ dimplications $A \rightsquigarrow B$ with $A, B \subseteq [w]$ having cardinalities $a = |A|$ and $b = |B|$. We record\footnote{This and all upcoming algorithms were programmed with MATHEMATICA. From among five random instances per quadtruple $(w, h, a, b)$ we recorded the two which realized the lowest and highest time. The time unit in Tables 8 to 10 is seconds unless stated otherwise.} the cardinality $|{\cal K}(\Theta)|$, the number of final $012e$-rows, and the running time.

\begin{tabular}{c|c|c|c}
$(w,h, a, b)$ & $|{\cal K}(\Theta)|$ & $012e$-rows & Time\\ \hline
 &    &  & \\
$(30, 50, 2, 8)$ & $916 647$ & $189$ & $0.3$\\
& $637 301$ & $722$ & $0.8$ \\ \hline
 &    &  & \\
$(30, 50, 8, 2)$ & $\approx 10^9$ & $146 382$ & $99$\\
& $\approx 10^9$ & $173 604$ & $123$ \\ \hline
 &    &  & \\
$(50, 1000, 2, 8)$ & $5326$ & $1932$ & $79$\\
& $5213$ & $2429$ & $103$ \\ \hline
 &    &  & \\
$(80, 50, 2, 8)$ & $\approx 5 \times 10^{19}$ & $28943$ & $43$\\
& $\approx 5 \times 10^{19}$ & $73390$ & $96$\\ \hline
 &    &  & \\
$(80, 20, 30, 8)$ & $\approx 10^{24}$ & $481 126$ & $258$\\
& $\approx 10^{24}$ &- & $> 15$ hrs \\ \hline  
 \end{tabular}
 
 Table 8: Knowledge Spaces ${\cal K}(\Theta)$ originating from random dimplication bases $\Theta$

Notice that the running time of our algorithm mainly depends on the {\it number of $012e$-rows} triggered by the input $\Theta$. Thus the sheer cardinality $|{\cal K}(\Theta)|$ is irrelevant, as e.g. witnessed by the $(80, 20, 30, 8)$ instance where $|{\cal K}(\Theta)| \approx 10^{24}$. Of course $10^{24}$ is way beyond the capacity of any algorithm generating ${\cal K}(\Theta )$ one-by-one. Estimating in advance the number $R$ of final $012e$-rows is  difficult but the following can be said. By the disjointness of rows we have $R \leq |{\cal K}(\Theta)|$ and thus $R \leq 2^w$, {\it independent} of the size of $\Theta$. To take an extreme case, for $w = 20$ even say $10^9$ dimplications will trigger at most $2^{20} \approx 10^6$ many $012e$-rows. The naive method of scanning all $2^{20}$ subsets $X$ of $[w]$ and testing $10^9$ conditions for each $X$ takes much longer even! For $w > 20$  a large number of dimplications  may still restrain ${\cal K}(\Theta)$ enough to keep $R$ at bay and to ridicule the naive method; see the $(50, 1000, 2, 8)$ instance. For a $(60, 1000, 2, 8 )$ instance this is less the case. To keep $w = 60$ while striving to have ${\cal K}(\Theta)$ and whence $R$ small, one would need to considerably increase $h = 1000$ (and with it Time). Of course not just $h$ but also the {\it type} of dimplication matters.

{\bf 7.2} For random families ${\cal B}$ of $n$ many $c$-element subsets of $[w]$ we calculate the corresponding base $\Theta_{PD}$ of prime dimplications (Theorem 1) and record both the cardinality $|\Theta_{PD}|$ and the computing time $T_1$. We replace $\Theta_{PD}$ by an equivalent base $\Theta_{\min}$ of minimum cardinality $|\Theta_{\min}|$ and record the time $T_2$ needed to do so. Using $\Theta_{\min}$ the Knowledge Space ${\cal K}({\cal B}) = {\cal K}(\Theta_{\min})$ gets calculated in time $T_3$. We further compare the total time $T = T_1 + T_2 + T_3$ with the time $T_{Dow}$ that Dowling's algorithm (see 3.2) takes to calculate ${\cal K}({\cal B})$.

Different to Table 8 in Table 9 the compression rate $|{\cal K}|$/(\# rows) is mostly low, and so Dowling's algorithm stands a chance, in fact is often superior. Specifically, for both the $(20, 50, 5)$ and $(20, 100, 5)$ instance in Table 9 we generated 5 random instances and picked the one for which $T/T_{Dow}$ was smallest and largest respectively. Notice the large gap between  $|\Theta_{PD}|$ and $|\Theta_{\min}|$ in the $(30, 20, 10)$ example. Calculating $\Theta_{PD}$ is comparatively swift\footnote{Among the many algorithms for hypergraph dualization [GV] the author coded a version of ``Berge multiplication'' with Mathematica.}, the bottleneck is the transition to $\Theta_{\min}$ (yet minimality is actually not essential, see the footnote in 8.2.2).  When one fixes $n$ and $c$ but lets $w \ra \infty$ then also $|{\cal K}| \ra \infty$. Numerical evidence indicates that the compression rate goes to $\infty$ as well (no formal proof is attempted).  Therefore Dowling's method looses out on the $e$-algorithm when $n, c$ are fixed and $w \ra \infty$. Of course in practice both become infeasible fast. For illustration consider the $(40, 100,2)$ instance.

\begin{tabular}{c|c|c|c|c|c|c|c|c} 
$(w, n, c)$ & $|\Theta_{PD}|$ & $|\Theta_{min}|$ & $|{\cal K}({\cal B})|$ & \# rows & $T_1$ & $T_2$ & $T_3$ & $T_{Dow}$\\ \hline
  &       &       &     &    & & & & \\ \hline
  $(20, 50, 5)$ & $6527$ & $2243$ & $68272$ & $15440$ & $2$ & $35$ & $562$ & $121$\\
  &  $6524$ & $2402$ & $67716$ & $18319$ & $3$ & $29$ & $868$ & $64$\\ \hline
  $(20, 100, 5)$ & $17530$ & $7127$ & $216 060$ & $40842$ & $24$ & $198$ & $4724$ & $1034$\\
  & $16743$ & $6784$ & $210815$ & $42362$ & $23$ & $180$ & $5818$ & $1044$\\ \hline
  $(30, 20, 10)$ & $26659$ & $843$ & $7168$ & $4267$ & $7$ & $1976$ & $103$ & 1\\
  $(30, 50, 10)$ & $215436$ & - & $175014$ & - & $376$ & - & - & $253$ \\ \hline
  $(40, 100, 2)$ & $40$ & $40$ & $228 \times 10^9$ & $10^6$ & 0 & 0 & $792$ & - \\ \hline
  \end{tabular}

Table 9: Knowledge Spaces ${\cal K}({\cal B})$ originating from random bases ${\cal B}$

{\bf 7.3} The meaning of generating Learning Spaces ``at random'' is not clear-cut. We proceeded as follows (although possibly in practice rather different types of Learning Spaces appear). For nonnegative integers $\mu, \lambda, \kappa$ start out with a $\lambda \mu$-element poset $(J, \leq)$ having $\lambda$ layers, each of cardinality $\mu$. For each poset element $x$, except for $x$ in the bottom layer, choose at random $\kappa$ lower covers in the layer below.

\begin{center}
\includegraphics[scale=0.6]{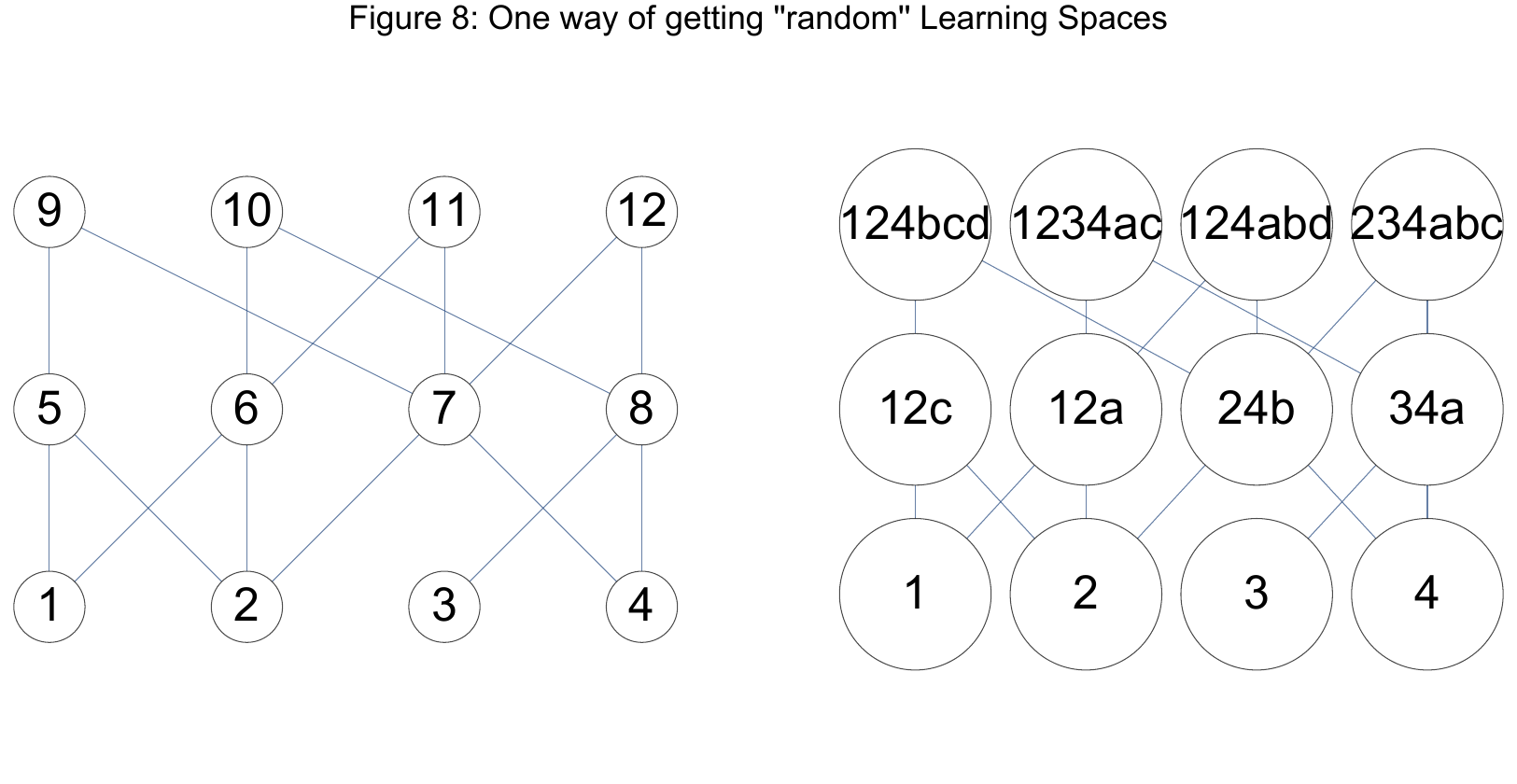}
\end{center}

For $(\mu, \lambda, \kappa) = (4, 3, 2)$ this is illustrated in Figure 8 on the left. Now recursively define $\lambda \mu$ many base sets $B_t$ as follows. Fix some set $NC$ (new colors) with $NC \cap \{1, \cdots, \lambda \mu \} = \emptyset$. Put $B_t  = \{t\}$ for $1 \leq t \leq \mu$. For $t = \mu +1, \mu+2, \ldots, \mu + \lambda$ let $B'_t$ be the union of all $\kappa$ many sets $B_s$ where $s$ ranges over the lower covers of $t$ in $(J, \leq)$. Put $B_t = B'_t \cup \{y\}$ where $y \in NC$ is chosen at random. To fix ideas, say $NC = \{a, b, c, d\}$. Then say\footnote{If not $a$ but again $c$ was chosen at random then $B_5 = B_6$, which is forbidden! Thus some supervision of the random process is required. That extends to the case that all $B_i$'s are distinct but $B_i \subseteq B_j$ for some $B_i, B_j$ on the {\it same level}. Then $B_i$ is discarded. Thus one may wind up with $|{\cal B}|$ smaller than $\lambda \mu$. Because the random process keeps on trying until each color in NC is picked at least once, we always have $|Q| = m+|NC|$.} $B_5 = B_1 \cup B_2 \cup \{c\} = \{1, 2, c\}$ and $B_6 = B_1 \cup B_2 \cup \{a\} = \{1, 2, a\}$. Similarly $B_7, B_8$ are obtained. Further say $B_9 = B_5 \cup B_7 \cup \{d\} = \{1, 2, 4, b, c, d\}$ and so forth (see Figure 8 on the right).

It is well known [FD] that $|{\cal B}({\cal K})| \geq |Q|$ for each Learning Space ${\cal K}\subseteq {\cal P}(Q)$ and that equality occurs iff ${\cal K}$ is distributive. In Table 10, we aim (as to why, see 7.4) to have ${\cal K}$ ``fairly distributive'', i.e. we choose NC as large as the random generation can exhaust it within reasonable time. Thus $|Q|= m+|NC|$ is fairly close to $|{\cal B}|$ throughout Table 10. As to the induced Learning Space ${\cal K}({\cal B})$, on the one hand it gets handled as in Table 9, i.e. we record its cardinality and the time $T_e$ for calculating $\Theta_{PD}$ and $\Theta_{\min}$, and running the $e$-algorithm on the latter. On the other hand ${\cal K}({\cal B})$ gets calculated as described in Section 6, i.e. $T_n$  denotes the total time to calculate $\Sigma_{JN}$ and $\Sigma_{\min}$, and to run the $n$-algorithm on the latter.

In the $(18, 22, 32, 36)$ instance we have $T_e = 3$ sec whereas $T_n = 18$ is behind. Incidentally both $\Theta_{\min}$ and $\Sigma_{\min}$ number to 26 (as indicated in brackets) but somehow $\Sigma_{\min}$ triggers more $012n$-rows than $\Theta_{\min}$ triggers $012e$-rows. Similarly for the $(18, 3, 2, 42, 54)$ instance. The tide turns in the other instances. They have in common that the base sets are larger, whence $\Theta_{PD}$ gets large (plausible from Theorem 1) and thus calculating $\Theta_{\min}$ is costly, and/or $\Theta_{\min}$ may itself remain fat.

\begin{tabular}{l|c|c|c|c|c} 
$(\mu, \lambda, \kappa, |Q|, |{\cal B}|$ & $|{\cal K}|$ & $e$-rows & $T_e$ & $n$-rows & $T_n$\\ \hline
 &   &   & & \\ \hline
$(18, 2, 2, 32, 36)$ & $4 \times 10^7$ & $4964 \ \ (26)$ & $3$ & $26768 \ \ (26)$ & $18$ \\ \hline
$(18, 3, 2, 42, 54)$ & $2 \times 10^8$ & $35711 \ (66)$ & $54$ & $349349 \ (64)$ & $492$ \\ \hline 
$(18, 2, 9, 32, 36)$ & $301308$ & $9205 \ \ (57)$ & $9$ & $3827 \ \ \ (26)$ & $3$ \\ \hline
$(50, 2, 25, 80, 100)$ & $10^{15}$ &  -\ \ \ \ \ $(3074)$ & - & $774680 \ (108)$ & $1685$ \\ \hline
$(5, 50, 2, 195, 213)$ & $1191092$ & $6195 \ \ (235)$ & $2446$ & $13431 \ \ (246)$ & $161$ \\ \hline
\end{tabular}

Table 10: Comparing the two ways ($012e$ or $012n$) to compress Learning Spaces

Thus in the $(50, 2, 25, 80, 100)$ instance the calculation of $\Theta_{\min}$ took only 23 seconds but $|\Theta_{\min}| = 3074$ was way too large for the $e$-algorithm to compete. It was aborted after 27 hours, having calculated $\approx 3 \times 10^{13}$ knowledge states out of $10^{15}$. The $(5, 50, 2, 195, 213)$ instance derives, in contrast to the others, from a thin and tall random poset. This forces the use of many distinct colors, and so {\it automatically} the Learning Space ${\cal K}$ becomes rather distributive. Here $T_e = 2446 = 29 + 2365 + 52$, where calculating the $16767$-element $\Theta_{PD}$ took a moderate $29$ sec, boiling it down to the $235$-element $\Theta_{\min}$ took a hefty $2365$ sec, and the actual $e$-algorithm finished in 52 sec. For comparison $T_n = 0 + 0 + 161$.

{\bf 7.4} If the Learning Spaces are ``less distributive'', i.e. feature fewer colors, then the cliques in the digraph $D({\cal L})$ become larger and with them $\Sigma_{JN}$, viewing that each $k$-clique $\{P_1, \cdots, P_k\}$ induces $k(k-1)$ implications in $\Sigma_{JN}$. The discussed standard implementation of the $n$-algorithm would thus potentially suffer from a large implication base $\Sigma_{\cal L}$, though not as badly as the $e$-algorithm can suffer from a large base $\Theta_{\min}$ of dimplications.
The good new is, the implications in $\Sigma_{JN}$ enjoy a lot of symmetry. In a nutshell, for fixed $P_j$ all $k-1$ implications induced by the arcs $P_i \mapsto P_j$ can be bundled to one ``compound''-implication whose algorithmic complexity isn't much higher than {\it one} ordinary implication. Furthermore all implications in $\Sigma_{po}$ have singleton premises and thus are benign. All of this gives rise to a special-purpose algorithm for implications such as the ones in $\Sigma ({\cal L})$ (work in progress).

\section{Two kinds of query learning: KST versus FCA}

Let us return to the question glimpsed at in Section 2, namely how to construct a Knowledge Space $(Q, {\cal K})$ by querying experts. Although I never published on query learning myself, I followed the development of Formal Concept Analysis (FCA) in this regard, from its beginnings in the 80's to the recent publication of [GO]. Most of [GO] is dedicated to {\it attribute exploration} which is the FCA term for query learning. 

What is the relation between FCA and Knowledge Space Theory (KST) anyway? Knowledge Spaces ${\cal K}$ are dual (w.r.t. $\cap, \cup$) to Concept Lattices, i.e. closure systems ${\cal C}$. So e.g. bases ${\cal B}({\cal K})$ correspond to formal contexts, and dimplications to implications. As an outsider to both ideologies, the author thinks the strong points of KST are the special features derived  for Learning Spaces, in particular the educational software company ALEKS based upon it [FD, p.10]. Vice versa, one strong point of FCA is attribute exploration. 

In Subsection 8.1 we scratch the surface of ``rejection-query-learning'' as it is developed in [GO] and improved upon in [RDB]. In 8.2 we turn to ``confirmation-query-learning'' as initiated in [K] and extended in [FD, 15.2].
In 8.3 attention gets restricted to Learning Spaces, with the effect that ideas get more crisp. The incorporation of our compression techniques follows in 8.4. In 8.5 we point out how both rejection-query-learning and confirmation-query-learning relate to the framework of Learning Boolean Functions.

{\bf 8.1} In a nutshell the query learning promoted in [Go] proceeds as follows. The exploration algorithm keeps on generating ``candidate'' implications $A \ra B$ (where $A, B \subseteq Q$ are sets of attributes) and each time asks the domain expert whether $A \ra B$ is a true implication in the domain to be explored. If the domain expert answers ``yes'' then the algorithm moves on to the next candidate implication. If she answers ``no'', she must provide a negative counterexample.

{\bf 8.1.1} For instance [GO,p.135], when exploring the structure of membership in various international organizations among the European countries, one candidate implication is $\{\mbox{Schengen area} \} \ra \{\mbox{Council of Europe}\}$. The answer is ``yes'' since each member of the Schengen area is member of the Council of Europe. Let $EU =$ European Union, $EUCU = EU$ Customs Union, $EEA =$ European Economic Area, and consider the candidate implication 
$$\{EUCU\} \ra \{EU, \mbox{Council of Europe}, EEA, \mbox{Eurozone}\}$$
This implication is false because a counter-example is e.g. provided by San Marino which belongs to $EUCU$ but not $EU$.

{\bf 8.1.2} Generally each counterexample provided by the expert is added as a $01$-row to a growing {\it context}. (In FCA a context is a $0,1$-table that encodes which ``objects'' have which attributes. Essentially the rows of the context match the meet irreducibles of the closure system ${\cal C}$ to be described.) The exploration algorithm stops based on some nice mathematics coupled to the implication base $\Sigma_{GD}$ alluded to in Section 5. Unfortunately this whole procedure of ``counterexample-based query learning'' takes as long as it takes to generate ${\cal C}$ one-by-one. This state of affairs is mitigated in [RDB] which ended decades of $\Sigma_{GD}$-fixation. Namely, for each $b \in Q$ all (or all important) prime implication $A \ra \{b\}$ of the current closure system (determined by the current context) get calculated in a way dual to Theorem 1. All of this still qualifies as counterexample-based query learning, but we rather call it {\it rejection-query-learning} to better fit terminology in 8.5.

{\bf 8.2} In [FD, 15.1.1] the following Naive Querying Algorithm (NQA) is proposed:

Step 1. \  Draw up the list of all the subsets $K$ of the domain $Q$.

Step 2. \ Successively submit all possible queries $(A, q)$ (meaning: does $A \rightsquigarrow \{q\}$ hold?).\\
\hspace*{1.35cm} Whenever the response from the expert(s) is positive, remove from the list of remaining \\
\hspace*{1.35cm} subsets all the sets $K$ disjoint from $A$ but containing $q$ (see (1)).

The comments on the NQA in [FD, 15.1.3] are quite harsh. In a nutshell:

Comment 1: Listing and processing all subsets $K \subseteq Q$ becomes infeasible as $Q$ grows large.

Comment 2: Worse, the list of queries $(A, q)$ is even larger [FD, p.302]. 

These ideas get refined in [FD, 15.2] but for many details the reader is referred to the original article of Koppen [K]. Both for NQA and the rejection-query-learning of 8.1 the exploration algorithm confronts the user with candidate formulas $\varphi$. For the former $\varphi$ is a dimplication $A \rightsquigarrow B$, for the latter an implication $\varphi = (A \ra B)$. However, the crucial difference isn't dimplication versus implication but is this. When $\varphi$ gets confirmed then NQA does work with $\varphi$ (in Step 2), whereas FCA does nothing. Vice versa, when $\varphi$ is rejected then NQA does nothing, whereas FCA must find a rejection ($=$counterexample) for $\varphi$. That's why we spoke of rejection-query-learning in 8.1 and now say that NQA is an example of {\it confirmation}-query-learning. Notice that both for NQA and FCA the current Knowledge Space ${\cal K}$, respectively closure space ${\cal C}$, {\it shrinks} whenever $\varphi$ has triggered work.

{\bf 8.3} Let us summarize from [FD, ch.16] how the NQA method criticized in Comments 1 and 2 is not so naive after all when it comes to {\it Learning Spaces}. The main idea is thus to approach the unknown target Learning Space $(Q, \ol{\cal K})$ by starting with the powerset ${\cal P}(Q)$ (or some other initial Learning Space) and by cutting off suitable\footnote{More details follow in 8.4.2.} chunks ${\cal D}_{\cal K}(A, q)$ of the current Learning Spaces $(Q, {\cal K})$ whenever a query $(A, q)$ gets answered in the {\it positive}. In formulas,

(21) \quad ${\cal K}$ gives way to ${\cal K} \setminus {\cal D}_{\cal K}(A, q)$.

Although ${\cal D}_{\cal K}(A, q)$ usually comprises {\it many} knowledge states (as opposed to 8.2), it would remain intractable to keep track of ${\cal K}$ by listing its members individually. The elegant solution in [FD, ch.16] is to keep track of ${\cal K}$ by merely holding on to its base ${\cal B}({\cal K})$. The algorithm stops when no more queries get answered positively. Then $\ol{\cal K}$ equals the last update ${\cal K}$, and can be generated by applying Dowling's algorithm to ${\cal B}(\ol{\cal K})$.

{\bf 8.4} Here come three ideas to improve upon 8.3, the third more speculative than the others.

{\bf 8.4.1} The target Learning Space $(Q, \ol{\cal K})$ may be too large to be generated from ${\cal B}(\ol{\cal K})$ by Dowling's algorithm. But we may get $\ol{\cal K}$ from ${\cal B}(\ol{\cal K})$ by either calculating a base $\Theta$ of dimplications and proceed as in Section 4, or by calculating a base $\Sigma$ of implications and proceed as in Section 6. 

{\bf 8.4.2} Even if the methods from Section 4 or 6 fail to deliver $\ol{\cal K}$ from ${\cal B}(\ol{\cal K})$ within reasonable time, we need not give up. To see why, let us unveil the structure of ${\cal D}_{\cal K}(A, q)$ in (21). It is
$${\cal D}_{\cal K}(A, q) := \{K \in {\cal K}| \, A \cap K = \emptyset \ \mbox{and} \ q \in K \} \ \ \mbox{[FD, p.337]}$$
Consider the following variant of the algorithm in 8.3. Apart from  ${\cal B}({\cal K})$ also keep track of ${\cal K}$, but in a compressed form as disjoint union of $012e$-rows, say ${\cal K} = r_1 \uplus r_1 \uplus r_2 \cdots \uplus r_s$. This format can be updated as follows. Putting ${\cal D} = {\cal D}_{\cal K}(A, q)$ one has
$${\cal K} \setminus {\cal D} \ = \  (r_1 \setminus {\cal D}) \uplus (r_2 \setminus {\cal D}) \uplus \cdots \uplus (r_s \setminus {\cal D}),$$
and so it suffices to show how $r_i \setminus {\cal D}$ can again be written as disjoint union of at most two $012e$-valued rows $r'$ and $r''$. First it is clear that $r= r_i$ satisfies
$$r \setminus {\cal D} \ = \ \{K \in r: \ q \not\in K\} \uplus \{K \in r: \ A \cap K \neq \emptyset \ \mbox{and} \ q \in K\}.$$

To fix ideas consider the $012$-row\footnote{Readers that mastered the technical details in Section 4 will have no problems extending it all to $012e$-rows.} $r$ in Table 11 and let $A = \{1, \cdots, 5\}, q = 6$. One verifies that indeed $r\setminus {\cal D} = r'\uplus r''$. This variant of the algorithm in 8.3 has the advantage that in the end (when no more positively answered queries occur) one has all these perks: The base ${\cal B}(\ol{\cal K})$, the Learning Space $\ol{\cal K}$ in compact format, and a base of dimplications of $\ol{\cal K}$ that matches the positively answered queries.

\begin{tabular}{l|c|c|c|c|c|c|c|c|c|} 
& 1 & 2 & 3 & 4 & 5 & 6 & 7 & 8 & 9 \\ \hline
&   &   &   &   &   &   &  &   & \\ \hline
$r$ & 2  & 0 & 2 & 2 & 0 & 2 & 1 & 0 & 2\\ \hline
$r'=$ & 2 & 0 & 2 & 2 & 0 & ${\bf 0}$ &  1& 0 & 2\\ \hline
$r'' =$ & $e$ & 0 & $e$ & $e$ & 0 & ${\bf 1}$ & 1 & 0 & 2 \\ \hline \end{tabular}

Table 11: The maneuver with ${\cal D}_{\cal K}(A, q)$

{\bf 8.4.3} Akin to the fact that bases of Learning Spaces ${\cal K}$ satisfy (LS), recall from (7a), (7b) that also PrimeDimp$({\cal K})$ has a very characteristic shape. One may thus ponder to approach $\ol{\cal K}$ by merely updating PrimeDimp$({\cal K}_i)$ for a sequence of Learning Space ${\cal K}_0 \supseteq {\cal K}_1 \supseteq \cdots \supseteq {\cal K}_s = \ol{\cal K}$. Presumably it will be helpful (or necessary) to update a representation of ${\cal K}_i$ by $012e$-rows as well. All details still need to be worked out. Suffice it to say that when $A \rightsquigarrow \{q\}$ is a prime dimplication for ${\cal K}_i$, it will stay a dimplication for ${\cal K}_{i+1}$ but possibly no longer prime.

{\bf 8.5.1} We draw on [CH1, chapter 7] which is a survey titled ``Learning Boolean functions with queries''. In brief, it is known in advance that the {\it target function} $f(x)$ to be learned belongs to a given class ${\cal C}$ of Boolean functions (such as monotone functions). The learner's objective is to identfy this function by asking questions ($=$ {\it queries}) about it. The most common kind of queries are {\it membership queries} and {\it equivalance queries}. The former asks for the function value $f(y)$ at a given ``candidate'' vector $y \in \{0,1\}^n$ specified by the learning algorithm. The expert's response to the query is $f(y)$, i.e. 0 or 1 whatever may be the case. In an equivalence query the learner asks whether his hypothesis function $h(x)$ (usually rendered by a Boolean formula) coincides with $f(x)$ for all $x \in \{0,1\}^n$. If the response is ``yes'', the learning process terminates. Otherwise the response is a {\it counterexample}, i.e. a vector $x$ with $h(x) \neq f(x)$. If $f(x) =1$, then $x$ is a {\it positive} counter-example, otherwise a {\it negative} one. According to a result of Angluin-Frazier-Pitt from 1992 the class ${\cal C}$ of all Horn functions is polynomial-time learnable in this manner [CH1, p.230].

{\bf 8.5.2} The paper [FP] propogates an alternative approach for learning Horn functions in that not candidate vectors $y$ but candidate {\it Horn clauses}\footnote{By definition a clause ($=$ disjunction of literals) is called {\it Horn} if it has at most one positive literal. If it has exactly one positive literal then it corresponds to what we called an implication $A \ra \{b\}$. Otherwise it is called an {\it impure} Horn clause. We also note that our term ``counterexample'' has two different meanings. In 8.5.1 it is a vector $x$, and in 8.5.2 an implication $A \ra \{b\}$.} are generated by the learning algorithm. Frazier and Pitt call their algorithm  {\it learning from entailment}. In this light the rejection-query-learning from above is a special case of learning from entailment where each Horn clause is an implication and where action is taken only upon negative counterexamples. And confirmation-query-learning is the special case where action is taken upon positive counterexamples.

{\it Acknowledgement:} I thank two References and the Acting Editor for their detailed constructive criticism. I am  grateful to Sergei Obiedkov for pointing out, late in the publication process, reference [YHM] which also proposes novel techniques for query learning in the framework of Learning Spaces. Although the compression issue is not adressed in [YHM], there are potential synergies to be explored in future work.

\section*{References}
\begin{enumerate}
\item[{[CH1]}] Y. Crama, P. Hammer, Boolean models and methods in mathematics, computer science, and engineering, Enc. Math. and Appl. 134, Cambridge 2010.
\item[{[CH2]}] Y. Crama, P.L. Hammer (eds.) Boolean functions, Enc. Math. and Appl. 142, Cambridge 2011.
\item[{[D]}] J.P. Doignon, Learning Spaces and how to build them, Lecture Notes in Computer Science 8478 pp.1-14 (2014).
\item[{[Di]}] B. Dietrich: A circuit set characterization of antimatroids, J. Combin. Theory B 43 (1987) 314-321.
\item[{[Dow]}] C.E. Dowling, On the irredundant construction of knowledge spaces. Journal of Math. Psych. 37 (1993) 49-62.
	\item[{[FD]}] J.C. Falmagne, J.P. Doignon, Learning Spaces, 417 pages, Springer 2011.
\item[{[FP]}] M. Frazier, L. Pitt, Learning from entailment: An application to propositional Horn sentences, Proceedings of the Tenth Internat. Conf. of Machine Learnig, 120-127, Morgan-Kaufmann 1993.
\item[{[GV]}] A. Gainer-Dewar, P. Vera-Licona, The minimal hitting set generation problem: algorithms and computation, arXiv:1601.02939v1.
\item[{[GO]}] B. Ganter, S. Obiedkov, Conceptual Exploration, Springer 2016.
\item[{[GKL]}] O. Goecke, B. Korte, L. Lovasz, Examples and algorithmic properties of greedoids, Lecture Notes in Mathematics 1403 (1986) 113-161.
\item[{[G]}] G. Gr\={a}tzer, Lattice Theory: Foundation, Birkh\"{a}user 2010.
	\item[{[JN]}] P. Janssen, L. Nourine, Minimum implicational basis for $\wedge$-semidistributive lattices, Inf. Proc. Letters 99 (2006) 199-202.
\item[{[K]}] M. Koppen, Extracting human expertise for constructing knowledge spaces: An Algorithm. Journal of Math. Psych. 37 (1993) 1-20.
\item[{[KLS]}] B. Korte, L. Lovasz, R. Schrader, Greedoids, Springer 1991.
\item[{[RDB]}] U. Ryssel, F. Distel, D. Borchmann, Fast algorithms for implication bases and attribute exploration using proper premises. Ann Math Artif Intell 70 (2014) 25-53.
	\item[{[W1]}] M. Wild, Compactly generating all satisfying truth assignments of a Horn formula, Journal on Satisfiability, Boolean Modeling and Computation 8 (2012) 63-82.
	\item[{[W2]}] M. Wild, The joy of implications, aka pure Horn formulas: mainly a survey. To appear in Theor. Comp. Science.
	\item[{[YHM]}] H. Yoshikawa, H. Hirai, K. Makino, A representation of antimatroids by Horn rules and its application to educational systems, arXiv: 1508.05465v1.
\end{enumerate}

\end{document}